\documentstyle[12pt,aaspp4,amssym]{article}
\input epsf
\def\Deg{${}^{\circ}$\llap{.}}
\def\Sec{${}^{\prime\prime}$\llap{.}}
\def\Min{${}^{\prime}$\llap{.}}
\def\mv{$M_{V}$(HB)~}
\lefthead{SANDQUIST ET AL.}
\righthead{CCD PHOTOMETRY IN M30}
\begin{document}
\title{Wide-Field CCD Photometry of the Globular Cluster M30.}

\author{Eric L. Sandquist\altaffilmark{1}, Michael Bolte,\altaffilmark{2}}
\affil{UCO/Lick Observatory, University of California, Santa Cruz,
CA~95064; erics@apollo.astro.nwu.edu, bolte@ucolick.org}
\author{G. E. Langer}
\affil{Department of Physics, Colorado College, Colorado Springs, CO
80903; elanger@cc.colorado.edu}
\author{James E. Hesser\altaffilmark{2}}
\affil{Dominion Astrophysical Observatory, Herzberg Institute of
Astrophysics, National Research Council of Canada, 5071 West Saanich Road,
Victoria, BC V8X 4M6, Canada; hesser@dao.nrc.ca}
\author{C. Mendes de Oliveira\altaffilmark{2}}
\affil{Instituto Astron\^omico e Geof\'{\i}sico (IAG), Av. Miguel
Stefano 4200 CEP: 04301-904 S\~ao Paulo, Brazil}
\altaffiltext{1}{Current address: Dearborn Observatory, Northwestern
University, 2131 Sheridan Road, Evanston, IL 60208}
\altaffiltext{2}{Visiting Astronomer, Cerro Tololo Inter-American
Observatory, National Optical Astronomy Observatories, operated by
AURA, Inc., under contract with the NSF.}

\dates

\begin{abstract}

We present new $VI$ photometry for the halo globular cluster M30 (NGC
7099 = C2137-174), and compute luminosity functions (LFs) in both
bands for samples of about 15,000 hydrogen-burning stars from near the
tip of the red giant branch (RGB) to over four magnitudes 
below the main-sequence (MS) turnoff. We confirm previously observed
features of the LF that are at odds with canonical theoretical predictions: an
excess of stars on subgiant branch (SGB) approximately 0.4 mag above
the turnoff and an excess number of RGB stars relative to MS stars. 

Based on subdwarfs with {\it Hipparcos}-measured parallaxes, we compute
apparent distance moduli of $(m-M)_{V} = 14.87\pm0.12$ and
$14.65\pm0.12$ for reddenings of E$(V-I)=0.06$ and 0.02 respectively.
The implied luminosity for the horizontal branch (HB) at these distances is
$M_V^{HB}=0.11$ and 0.37 mag.  The two helium indicators
we have been able to measure ($R$ and $\Delta$) both indicate that
M30's helium content is high relative to other clusters of similar
metallicity. M30 has a larger value for the parameter  $\Delta
V_{TO}^{HB}$ than any of the other similarly metal-poor clusters for
which this quantity can be reliably measured.  This suggests that M30
has either a larger age or higher helium content than all of the other
metal-poor clusters examined. The color-difference method for measuring
relative ages indicates that M30 is coeval with the metal-poor clusters
M68 and M92.
\end{abstract}


\section{Introduction}\label{intro}

This paper is the second in a series investigating the evolved-star
populations in nearby globular clusters. With the large-field CCD
imagers now available it is possible to measure nearly complete
samples of giant stars in clusters, and at the same time measure stars
faint enough that we can normalize the luminosity functions (LFs) to
the unevolved main sequence. Because the LFs for evolved stars
directly probe the timescales and fuel consumed in the different
phases of stellar evolution, they provide a stringent test of the
models for the evolution of low-mass stars.  These models are the
basis for our use of globular clusters to set a lower limit to the age
of the Universe and are a fundamental tool in the interpretation of
the integrated spectra and colors of elliptical galaxies.

The subject of this study is M30 (NGC 7099 = C2137-174), a relatively
nearby cluster ($\sim 7$ kpc; Peterson 1993) at high galactic latitude
($b$ = $-$46\Deg8). M30 has a high central density ($\log \
(\rho_{0} / (M_{\odot} / \mbox{pc}^{3})) = 5.9$), a moderate total
mass ($\log (M / M_{\odot}) = 5.3$; Pryor \& Meylan 1993), and is at
the metal-poor end of the cluster [Fe/H] distribution. It is one of
approximately 10\% of clusters that have cusps at the core of
their surface brightness profiles, and it also has one of the largest
radial color gradients of any cluster (Stetson 1991b).

Previous studies of the LF for stars in metal-poor clusters have
uncovered unexpected features. In a LF formed from the
combination of CCD-based observations of the clusters M68 (NGC 4590 =
C1236-264), NGC 6397 (C1736-536), and M92 (NGC 6341 = C1715+432),
Stetson (1991a) found an excess of stars on the subgiant branch (SGB)
just above the main-sequence turnoff (MSTO).  Bolte (1994) and
Bergbusch (1996) both observed M30 using a mosaic of small-field CCD
images and found an excess of SGB stars. (The SGB is defined here as
the transitional region between the main-sequence turnoff and the base
of the red giant branch at the point of maximum curvature.)

Another unexpected observation involving LFs is a mismatch between
theoretical predictions and the observed size of the ``jump'' dividing
the main sequence (MS) and the red giant branch (RGB). When normalized
to the MS, there is an excess of observed RGB stars compared to models
(Stetson 1991a, Bergbusch \& VandenBerg 1992, Bolte 1994, Bergbusch
1996), although this has been disputed by Degl'Innocenti, Weiss, \&
Leone (1997). These results might be explained by the action of core
rotation (VandenBerg, Larson, \& DePropris 1998), or perhaps (as
discussed later) we are witnessing the results of deep mixing and the
delivery of fresh fuel into the hydrogen shell-burning regions. Langer
\& Hoffman (1995) suggested that, if the abundance patterns of light
elements seen in bright cluster giants (e.g. Kraft 1994) are due to
deep mixing, hydrogen-rich envelope material is almost certainly mixed
into the hydrogen burning shell (prolonging the giant phase of
evolution), and some of the helium produced is returned to the
envelope.  Because of the potential importance of such non-standard
physics in stars, and because of the caveats associated with earlier
LF studies, the most productive next step is to derive better LFs in a
number of Galactic globular clusters (GGCs).

In the next section, we describe our observations of the cluster.
In \S 3, we discuss the features observed in the color-magnitude
diagram, describe the method of computing the luminosity functions,
and present the results of artificial star experiments. In \S
4, we discuss the constraints that can be put on the global parameters
of the cluster --- metallicity, distance, and age. The method of data
reduction is described in Appendix A.

\section{Observations}

The data used in deriving the $V$- and $I$-band LFs of M30 were taken
on July 7/8, 1994 at the Cerro Tololo Inter-American Observatory
(CTIO) 4 m telescope.  In all, six exposures of 120 s, one exposure of
60 s and two exposures of 10 s were made in $V$, and six exposures of
120 s, one exposure of 60 s, and one exposure of 10 s were made in
$I$.  All frames were taken using the 2048 $\times$ 2048 pixel ``Tek
\#4'' CCD chip, which has a sampling of about 0\Sec44 per pixel, and a
field 15$^{\prime}$ on a side. These exposures were reduced
individually for the purpose of constructing the color-magnitude
diagram. In performing artificial star experiments and deriving
the LF, the three best-seeing images in both $V$- and $I$-bands were
combined into master long-exposure images. The frames were
centered approximately 2$^{\prime}$ east of the cluster center, in
order to avoid a bright field star nearby.

The night of the 4 m observations was not photometric. In order to set
the observations on a standard photometric system, we used
observations made at the CTIO 1.5~m telescope on one photometric night
(October 18/19, 1996). The detector used was the ``Tek \#5'' 2048
$\times$ 2048 CCD, having a field of about 14\Min8 on a side. Landolt
(1992) standard star observations were used to calibrate a secondary
field that overlapped the 4 m field. On that night, 10~s and 120~s
exposures were taken in each band, along with exposures of 27
standards in 7 Landolt fields. A sample of 118 stars having $12.9 < V
< 1.5$ and $-0.03 < (V - I) < 1.31$ was calibrated as secondary
standards in this way. The field was centered approximately
$5^{\prime}$ south of the cluster center.

During the same run on the 1.5~m telescope, frames were taken of M30
on the non-photometric night of October 16/17. Five additional
exposures were taken in each band (20 s, 200s, and 3$\times$600s in V,
and 15 s, 180 s, and 3$\times$600s in $I$). The details of the data
reduction and calibration are described in Appendix A.

\section{The Color-Magnitude Diagram (CMD) and Luminosity Functions (LFs)}

\subsection{The CMD}

In Figure~\ref{m30cmd}, we plot the total $VI$ sample of 25279 stars
(upper panel) and a sample that has been restricted in projected
radius to $110^{\prime\prime} < r < 10^{\prime}$ from the cluster
center (lower panel).  The inner radius was chosen in accord with the
restriction placed on stars to be used later in the LFs, while the
outer radius restriction was chosen to exclude regions that were
affected by field star contamination.

Fiducial points for the MS and lower RGB of the clusters were
determined by finding the mode of the color distribution of the points
in magnitude bins.  The fiducial line on the upper RGB was determined by
finding the mean color of the stars in magnitude bins. Once a mean was
determined, stars falling more than $7\sigma$ from the fiducial point
were discarded (so as to eliminate AGB and HB stars, as well as blends
and poorly measured stars), and the mean redetermined. This procedure
was iterated until the star list did not change between iterations.
At the tip of the RGB and on the AGB, the positions of individual
stars were included as fiducial points if they appeared to be
continuations of the mean fiducial line. The fiducial line for the HB
was obtained by determining mean points in magnitude bins for the blue
tail, and in color bins for the horizontal part of the branch.  No
smoothing has been applied. Table~\ref{fidtab} lists the fiducial lines for our
samples, as well as the number of stars used in computing each point.

\subsection{The LFs and Incompleteness}

The procedure used to correct the ``observed'' LF back to the ``true''
LF is described in detail in Sandquist et al. (1996). As in that
paper, we carried out artificial star tests on only four frames: a
long exposure frame (composed of the average of the three best-seeing
images) and a short exposure frame in both $V$ and $I$ band. In 11
runs, 20965 artificial stars were processed.

ALLFRAME's coordinate transformations were found to
be unable to follow the nonuniform spatial distortions introduced by
the 4~m field corrector. To avoid this problem, we reduced all of the
frames through ALLSTAR as usual, derived a master detected star
list for each filter, and rereduced the frames in ALLSTAR with the
improved positions. This procedure improved the overall
quality of the photometry (as judged by the scatter around the
fiducial lines of the cluster), as ALLFRAME normally does.

In Figures~\ref{delta} -- \ref{fsfig} we plot our computed values for
median magnitude biases $\delta_{V}$, median external error
$\sigma_{ext}(V)$, and completeness probability $f(V)$ as a function
of magnitude and radius. There is little variation in most of the
quantities until the innermost radial region ($r < $ 2\Min0, where
crowding of stellar images is worst) is reached.

One change we have made since our first study was in the error
estimation. The uncertainty in the incompleteness factor $f$ was
previously found by simultaneously varying $F, \sigma$, and $\delta$
in such a way as to cause the maximum change in $f$ away from our best
value. The magnitude of this change was used as the error estimate. We
have improved this, following a suggestion by Bergbusch (1996),
by estimating the error by varying $F, \sigma,$ and $\delta$ individually
and adding the resulting error estimates in quadrature.

Using this information, we eliminated stars from consideration for the
LF if they fell far enough away from the nearest point on
the fiducial line of the cluster. The ``distance'' was defined in
terms of difference in magnitude and color divided by their respective
external errors, and then added in quadrature. So, stars were
eliminated if $(\Delta V / \sigma_{ext}(V))^2 + (\Delta (V-I) /
\sigma_{ext}(V-I))^2 > 25$. On the upper RGB, where
contamination by the AGB could be a factor, we adjusted the error
cutoff by hand until we were sure the AGB stars were being eliminated,
but not at the expense of the RGB stars.

The luminosity functions are listed in
Tables~\ref{vlftab} and \ref{ilftab}. Totals of 14772 and 14507 stars
were used in creating the $V$- and $I$-band LFs.  Stars on the MS and
SGB were only included if they fell more than 2\Min0 from the center
of the cluster. This reduced the significance of crowding
effects on the photometry.  Stars with magnitudes $V < 16.9$ (or $I <
17.0$) were included in the determination of the LF to within
$30^{\prime\prime}$ of the center of the cluster.  This was done to
get a better indication of the ``global'' LF for the red giants since
mass segregation is expected to occur in M30, which would affect the
RGB star counts taken from a small range of radii.

\section{Discussion}

\subsection{Reddening and Metallicity}\label{red}

Most previous studies have made estimates of E($B-V$). We will assume
E$(V-I)=1.25 \mbox{E}(B-V)$ (Cardelli, Clayton, \& Mathis 1989) for
the rest of the discussion and refer to the reddening in E$(V-I)$.
M30 is situated well out of the Galactic plane ($b^{II}= -47$) and the
reddening is likely small (E$(V-I) < 0.1$).  Reddening maps of
Burstein \& Heiles (1982) indicate that E$(V-I) < 0.04$. Reed, Hesser
\& Shawl (1988) derive a {\it negative} reddening based on the
comparison of M30's integrated color and spectral type. Zinn (1980)
gets a value of 0.01 from integrated-light measurements. Reed, Hesser,
\& Shawl's data indicates that M30 appears to have the bluest
intrinsic colors of all of the globular clusters they examined.
However, M30 is the cluster that shows the strongest color gradient of
any GGC and reddening values measured from integrated colors or
spectra will depend on the range of radii over which the observations
are made. Given the sense of the gradient, bluer towards the center,
it seems likely that these reddening measurements will be biased to
lower values.
  
Dickens (1972) and Richer, Fahlman, \& VandenBerg (1988; hereafter
RFV) independently derived significantly higher values, E($V-I)\sim
0.08$, based on the UBV colors of blue HB stars in M30. On the other
hand, $U$-band photometry is notoriously difficult to calibrate and
the precision of reddening estimates from color-color plots is $\sim
0.04$ mag even in the best cases. Neither the Dickens nor the RFV
study appears to have had a photometric calibration good enough to
warrant error bars less than this. Differential CMD comparisons with
M92 (Vandenberg, Bolte, \& Stetson 1990; hereafter VBS) and M68
(Figure~\ref{m30vsm68}) imply E$(V-I)\sim 0.05$ to 0.06. The recent
reddening maps based on IRAS and COBE measurements of far-IR flux from
infrared cirrus suggest E($V-I)\sim 0.063$.

We can make our own estimate using using
Sarajedini's (1994) simultaneous reddening and metallicity method (for
$V_{HB} = 15.04\pm0.08$ and $(V-I)_{g} = 0.97\pm0.02$, where the
quoted errors allow some room for calibration errors). We find E$(V-I)
= 0.065\pm0.003$.  The errors were derived
from Monte Carlo tests with the quoted errors on $V_{HB}$ and
$(V-I)_{g}$.  Most of the reddening estimates we have thus far
indicate a relatively high value E$(V-I) = 0.06\pm0.02$.

The compilation of Zinn \& West (1984) has [Fe/H] $= -2.13
\pm 0.13$ and numerous studies since have determined values
between $-1.9$ and $-2.3$ (Geisler, Minniti, \& Clari\'{a} 1992;
Minniti et al. 1993; Carretta \& Gratton 1996). From the simultaneous
reddening and metallicity method above, we find [Fe/H]$=-2.01\pm0.09$.

Anticipating later discussion of the level of the HB, we
examine the effects of an anomalously high \mv on the
simultaneous reddening and metallicity method. A high \mv value can
come about due to a high helium abundance, whether primordial or the
result of a ``deep mixing'' scenario (Langer \& Hoffman 1995). If the
``true'' \mv value (in the absence of helium enrichment) is fainter by
0.10 mag, then we would have $V_{HB} = 15.14$, and calculate [Fe/H] $=
-2.11\pm0.07$ and E$(V-I) = 0.068\pm0.002$. 

\subsection{Distance Modulus}\label{dist}

Recently Reid (1997), Gratton et al. (1997) and Pont et al. (1998)
have used subdwarf parallaxes measured with the {\it Hipparcos}
satellite (ESA 1997) to redetermine the distance moduli to several of
the best observed clusters.  The general result of these studies is to
increase cluster distance moduli by 0.2 to 0.4 mag, implying high
luminosities for the horizontal branches (as bright as M$_V\sim 0.1$
mag for the [Fe/H]$\sim -2$ clusters). For M30 specifically, Gratton
et al. find (m-M)$=14.96\pm 0.10$ (for E$(V-I)=0.05$) although this
is based on only three subdwarfs. We repeat this exercise for M30 with our
data and a larger set of subdwarfs.
This method is sensitive to uncertainties in the color of the unevolved
main-sequence, which result from zero-point errors in the photometric
calibration, the reddening uncertainties already mentioned, and
uncertainties in the placement of the main-sequence fiducial line.  Our
data for M30 are not optimum for using subdwarf fitting to measure the
distance, primarily because of the uncertainty in the reddening, but
also because we suffer from each of the other problems to a small degree.

Nevertheless, we selected subdwarfs that satisfy the following
criteria: parallaxes from the {\it Hipparcos} mission having relative
errors $\sigma_{\pi} / \pi < 0.15$, metal abundances from the study of
Gratton, Carretta, \& Castelli (1996), and $VI$ photometry.  The
restriction on parallax error was chosen to minimize the effect of bias
corrections. (As a result, Lutz-Kelker corrections only change our
derived distance moduli by 0.01 mag.) The Gratton et al.  metallicity
scale was chosen for its homogeneity and because it minimizes
the possibility of systematic abundance errors with respect to Carretta
\& Gratton globular cluster metallicities. When available, we used $VI$
photometry for the subdwarfs tabulated in Mandushev et al.  (1996). In
the remaining cases, we followed their procedure of combining
literature values from the following sources: Carney \& Aaronson
(1979), Carney (1980, 1983b), and Ryan (1989, 1992). The studies
involving Carney all used the Johnson $I$ filter, so we applied the
transformation from Carney (1983a) to convert them to the Cousins
system.  Known spectroscopic binaries were excluded, and, for the cases
where it has been measured, any reddening of the subdwarfs (at most a
very small amount) has been subtracted.  Our sample of subdwarfs and
metal-poor subgiants is shown in Table~\ref{subdtab}.  We used the
subdwarfs to estimate the distance to M30 in two different ways.

First, to simultaneously 
estimate the distance and E$(V-I)$ of M30 we created
a grid of chi-square-like sums. The M30 main sequence between
$19<V<22.8$ was represented by a third-order polynomial. 
For ranges of m-M and E$(V-I$), the minimum distance 
of each subdwarf from the main-sequence polynomial 
was calculated. This distance was normalized by the combined
errors in the subdwarfs' colors and magnitudes and the 
main-sequence fiducial uncertainties in color and magnitude
(assumed to be 0.04 mag in color and 0.05 mag in $V$ magnitude).
Our ``$\chi^2$'' sum is:

$$\chi^2= \sum_i{\frac{(V_i - V_{\rm fiducial})^2 + ((V-I)_i - (V-I)_{\rm fiducial})^2}
{\sigma^2}}$$ 

We will refer to this as the chi-square sum although it does not
match the usual definition of chi-square and it is not normalized
in a way to give true confidence intervals. Because the slope of the
main-sequence changes with $V$, there is a different weighting
given to $\Delta V$ and $\Delta (V-I$) for each subdwarf.
The minimum chi-square values are for
m$-$M$\sim 14.6$ and E$(V-I)\sim 0.02$, although the reddening in
particular is poorly constrained.

Our second approach was to fit the M30 main sequence to the
subdwarfs using only the distance modulus as a variable
for two E($V-I$) values: 0.02 and 0.06.
Table~\ref{dmtab} shows the changes in this
value for different subsets of our sample and for different input
data.  (The quoted errors include contributions from the scatter of
values in the subdwarf fit, a cluster reddening error of 0.02 mag, and
the absolute cluster metallicity error of 0.2 dex.) Clearly, reddening
uncertainty is dominant in the total uncertainty.  Two fits at
different reddening are shown in Figure~\ref{subdvred}. If we use the
value of metal content given by Zinn \& West ([Fe/H] $= -2.13$) along
with the Carney et al. (1994) abundance scale for the subdwarfs (which
has roughly the same zero point as Zinn \& West), the distance modulus
is increased by only a few hundredths of a magnitude. Restricting the
sample to only metal-poor ([Fe/H]$< -1.3$) subdwarfs also does not
significantly change the distance modulus.

Considering the number of distance modulus measurements for M30 in the
literature, it is best to try to compare using a common reddening
of E$(V-I)=0.02$ (and using $\partial (m-M)_{V} / \partial
\mbox{E}(V-I) \approx 6$ to correct values). From the two methods
we presented here we have 14.6 and 14.65. From the pre-Hipparcos
ground-based measurements of Bolte (1987) and RFV, we have 14.62 and
14.68 (the RFV value must also be corrected for their lower assumed
metallicity for M30).  For the Hipparcos-based distances of Reid
(1997) and Gratton et al.  (1997), we find 14.75 and 14.87 (although
Gratton et al. use only three subdwarfs in their fit). It is clear that
our distance moduli are consistent with the ground-based measurements,
and over 0.1 mag smaller than the other Hipparcos-based measurements.
Neverthless, the different distance measures are in good general
agreement, at least for a fixed reddening value.
Although it is not crucial for the conclusions
that follow, we will adopt (m-M)$_V = 14.7$ for the rest of the paper.
From Figure~\ref{subdvred}, it is clear that the smaller reddening and
distance modulus values are more consistent with the model
isochrones and, for our preferred larger reddening
value of E$(V-I)=0.06$, the models do not match the
shape of the M30 fiducial above the turnoff.

\subsection{The $R$ Method}\label{bright}

The ratio $R = N_{HB} / N_{RGB}$, where $N_{RGB}$ is the number of RGB
stars brighter than the luminosity level of the HB, is the traditional
quantity used to estimate the helium abundance of stars in globular
clusters. Dickens (1972) first noted that M30's value for
``R'' was unusually large and Alcaino \& Wamsteker (1982) claimed a
significant gradient in this ratio in the sense of a small
value at the center of the cluster increasing to one of the
largest measured in any cluster at large radii.
We have sufficient numbers of stars on the red and blue
sides of the instability strip to define the level of the HB.  From
five stars near the blue edge of the instability strip (with colors
$0.25 < (V-I) < 0.35$), we calculate an average magnitude $V_{HB,b} =
15.08 \pm 0.06$. From six stars on the red side of the instability
strip, we find $V_{HB,r} = 14.90 \pm 0.06$. All stars used in these
averages were found at radii greater than 1\Min0 from the center of
the cluster, so that the photometry should be quite accurate.

Because the estimate of the HB magnitude is crucial in later
arguments, we examine the topic further. Our blue side estimate
is consistent with those of Bolte (1987) and Bergbusch (1996), even
with the different calibrations of our respective datasets. Because it
is possible that the red HB stars in M30 are evolved, it is wise to
check this possibility before simply interpolating between the two
sides. The number of stars (8 with projected radius $r >$ 1\Min0,
compared to 93 on the blue HB) at $V \sim 14.9$ is roughly consistent
with timescales for stars evolving from the blue HB, as the
evolutionary tracks tend to parallel each other closely through this
part of the CMD as they move toward the AGB.  Thus, we have chosen to
look at other clusters of similar metallicity with large, well-studied
RR Lyrae populations to compute a magnitude correction to go from the
red edge of the blue HB into the middle of the instability strip.  For the
clusters M15 (Bingham et al. 1984), M68 (Walker 1994), and M92 (Carney
et al. 1992) we find agreement on a correction of $-0.04$ mag. Thus, we
have $V_{HB} = 15.04 \pm 0.06$.

To define the RGB star sample for the $R$ method, we need to establish
the relative bolometric magnitudes of the HB and RGB stars. We have
used the HB models of Dorman (1992) in conjunction with the isochrones
of Bergbusch \& VandenBerg (1992; hereafter BV92). The stellar models
involved in these studies were computed with a consistent set of
physics and compositions.  Although the composition used is somewhat
out-of-date, the {\it differential} bolometric corrections should be
fine. The corrections as a function of [Fe/H] can be approximated by:
\[ \Delta V_{BC} \equiv V_{RGB} - V_{HB} = 0.709 + 0.548 \mbox{[M/H]} +
0.229 \mbox{[M/H]}^{2} + 0.034 \mbox{[M/H]}^{3} .\] Because
$\alpha$-element enhancements influence the position of the HB and RGB
in the CMD like a change in [Fe/H] (Salaris, Chieffi, \& Straniero
1993), they must be taken into account when computing [M/H].

For M30, we find that $\Delta V_{BC} = 0.27$. We have used [M/H]
$=-1.70$ (correcting [Fe/H] by 0.21 dex for [$\alpha$/Fe]$=+0.3$),
although for this range of metallicities, changes in metallicity have
a very small effect on $\Delta V_{BC}$. Because of contamination and
blending problems toward the center of the cluster, we restrict our
samples of RGB and HB stars to $r > 20^{\prime\prime}$.  With this
choice, we find $R = 1.45\pm0.20$.  This makes M30's $R$ the
highest of any of the clusters examined. In Table~\ref{clusttab}, we
present $R$ values that have been derived from published photometry
for the clusters similar in [Fe/H] to M30 -- 
M68 (Walker 1994), M53 (Cuffey 1965), NGC 5053 (Sarajedini \&
Milone 1995), NGC 5466 (Buonanno et al. 1984), and M15 (Buonanno et
al.  1983) -- along with several more metal-rich clusters.

Photometry exists for the central $25^{\prime\prime}$ of M30 from
the Hubble Space Telescope (Yanny et al. 1994; hereafter YGSB). By
merging their list with ours and eliminating common stars, we have created a
master list of HB and RGB stars that completely covers
the cluster out to 7\Min0, with portions included out
to about $12^{\prime}$. The data for the full sample is presented in
Table~\ref{poptab}. Using this sample, we find a global value
for $R$ of $1.49\pm0.18$. (Even if our value of $V_{HB}$ is too
bright, and if we use the magnitude of the red edge of the blue HB, we
find $R = 1.36\pm0.21$ --- still a high value in a relative sense.) 

This global $R$ value for M30 is on firm ground, because there is no
place for the bright stars to hide. The photometry is easily good
enough to distinguish between the AGB and RGB star samples, so this is
not a source of uncertainty. (In fact, M30 may be deficient in AGB
stars as well as RGB stars.) The use of the lower Buzzoni et al.
(1983) value $\Delta V_{BC} = 0.15$ for the differential bolometric
correction would increase the high $R$ value.
Approximately 30 additional RGB stars would have to be included to
bring M30's value in line with that of other clusters --- a
24\% change in the sample size.

There are a few potential explanations for a {\it global} depletion of
bright giants in this cluster.  First, because the ratio $R$ is a
helium abundance indicator, the abnormally high value could indicate a
higher-than-average helium abundance in M30 stars more luminous than
the HB. This does not necessarily imply high Y for lower luminosity
stars, since a deep-mixing mechanism would also be expected to dredge
up freshly produced helium (Sweigart 1997). Second, the environment
within the cluster may affect the stellar populations by a mechanism
that truncates RGB evolution and/or produces additional HB stars. We
now consider the arguments for the two sides.

\subsubsection{Helium Abundance}\label{heabund}

Using the Buzzoni et al. (1983) calibration and our M30 $R$ value, we
find $Y = 0.24 \pm 0.02$. The value derived using $V_{HB}$ of the blue
edge of the instability strip gives $Y=0.23\pm0.02$. That $V_{HB}$ is
definitely a faint limit, making $Y=0.23$ a lower limit. In any case,
the $R$ value for M30 is significantly higher than those for other
clusters when the revised differential bolometric corrections are
used. [The low value of $Y(R)$ for the other clusters is discussed in
Sandquist (1998).]

One check we can make is to examine other helium
indicators to see if they also indicate a high abundance relative to
other clusters.  Caputo, Cayrel, \& Cayrel de Strobel (1983)
introduced two indicators: $A$ (the mass-luminosity relation for RR
Lyrae stars of ab type) and $\Delta$ (the magnitude difference between
the HB and the point on the MS where the dereddened $B-V$ color is
0.7). While M30's RRab variables have not been studied to the extent
necessary to compute $A$, we can compute a $\Delta$ value
from the $BV$ photometry of Bolte (1987) and RFV.
Assuming for both that E$(V-I)=0.06\pm0.02$, we find $6.42\pm0.12$ and
$6.28\pm0.13$ respectively, where the primary contributions to the
error are the uncertainty in the reddening and the small number of
stars used to define the HB magnitude. (The lower RFV value can be
traced to a fainter HB magnitude relative to Bolte's data.)  We have
computed comparison values for the clusters M68 (McClure et al.
1987, Walker 1994), and M15 (Durrell \& Harris 1993), as summarized in
Table~\ref{clusttab}. The $\Delta$ values for the other clusters agree
with the theoretical value of 6.30 for [M/H] $= -1.82$ ([Fe/H]$=-2.03$).

We can directly compare our data with $VI$ for other clusters if we
redefine $\Delta$ (hereafter, $\Delta_{B-V}$) by choosing the MS point
to have $(V-I)_{0} = 0.85$.  From isochrones this is approximately
equivalent to $(B-V)_{0} = 0.7$.  From our fiducial line, we find
$\Delta_{V-I} = 6.36\pm0.10$ for E$(V-I)=0.06\pm0.02$. $VI$ data
exist for M92 (Johnson \& Bolte 1998) and M68 (Walker 1994). We find
that $\Delta_{V-I} = 6.25\pm0.12$ (E$(V-I)=0.025\pm0.0125$) and
$6.11\pm0.07$ (E$(V-I)=0.0875\pm0.0125$), respectively. The M92 value
relies essentially on one star for the HB magnitude, so the
$\Delta_{V-I}$ value is uncertain.

For $\Delta_{B-V}$ and $\Delta_{V-I}$ we see evidence (though not
overwhelming) that the M30 value is high compared to other clusters.
The values above indicate that M30's helium abundance is high by about
0.02 for [M/H]$=-1.7$ if the helium is primordial, or 0.03 if the
helium enrichment only affects the level of the HB, as in the deep
mixing scenario. (Note that a lower value for the reddening would bring
the $\Delta$ values into consistency with the other clusters.) A high
helium abundance would tend to make the HB distribution bluer on the
whole.  According to Fusi Pecci et al. (1993), the color of the peak of
the HB star distribution in M30 is one of the bluest, but other
clusters are rather close (M53 is bluer, M15 has approximately the same
peak color, and M92 and NGC 5466 are slightly redder).

\subsubsection{Environmental Effects}

Because M30 has a core of high stellar density, we consider the
possibility that this environment has influenced the populations of
evolved stars in the cluster. M30 has one of the most
robustly determined color gradients (approximately linear in $\log r$:
$\sim +0.20$ mag dex$^{-1}$; Piotto, King, \& Djorgovski 1988) of all
the globular clusters in the Galaxy. The sense of the color gradient
is such that the integrated colors become bluer towards the cluster
center.  In M30 and other post-core-collapse clusters it has been
suggested that the color gradient is due to a decrease in the ratio of
RGB-to-BHB stars resulting from stellar interactions in the dense
cluster cores.  Although Djorgovski \& Piotto (1993) claim the color
gradient measured in M30 is due to a deficit of RGB stars in the inner
few tens of arcseconds, Burgarella \& Buat (1996) show that the
gradient (which extends to radii $>2^\prime$) is not due to
differences in the spatial distribution in the evolved-star
populations or in the blue stragglers. Our results are consistent with
this latter claim.

Buonanno et al. (1988) claimed to have detected a radial variation in
the ratio $R = N_{HB}/N_{RGB}$ on the basis of a smaller sample of
stars. From our sample, we find $R$ ranges from about $1.30\pm0.26$ in
the inner $30^{\prime\prime}$ to $1.35\pm0.27$ for
$30^{\prime\prime}<r< 100^{\prime\prime}$ to $1.83\pm0.36$ for stars
with projected radius $r > 100^{\prime\prime}$.  There may be marginal
evidence for a difference between the outer annulus, and the inner
two, but over the range of radii for which the bluer-inward color
gradient has been observed, there is no evidence of a trend in the
bright populations. The sense of the difference between the outer and
inner populations is in any case opposite to that required to make the
color gradient.  The cumulative radial distribution (Figure~\ref{crd})
also shows no strong trends in radius, contrary to claims in other
studies with smaller samples (Buonanno et al. 1988, Piotto et al.
1988), but in agreement with studies of the core (Yanny et al. 1994,
Burgarella \& Buat 1996).  A Kolmogorov-Smirnoff test indicates a 33\%
chance that the two samples are drawn from the same distribution. The
inhomogeneity of the RGB sample seems to be responsible for this
noncommittal probability.

We conclude that despite the color gradient in M30, and the apparently
ripe conditions for interactions to alter the stellar populations, 
environment-based processes are not responsible for the high R
value we measure.

\subsection{Age}

\subsubsection{Relative Age Indicators: $\Delta V_{TO}^{HB}$ and
$\Delta (V-I)$}\label{sage}

Based on the color-difference method, VBS claimed that the most
metal-poor clusters, including M30, are coeval at the level of 1 Gyr.
Our comparison (Figure~\ref{m30vsm68}) of the fiducial lines
of M30 and M68 (Walker 1994) in the neighborhood of the SGB indicates
that the ages of M30 and M68 (assuming similar main-sequence
Y and [$\alpha$/Fe]) are nearly identical.  With our uniform
calibration of MSTO and evolved stars in M30, we can determine with
good precision the other commonly applied age estimator $\Delta
V_{TO}^{HB}$.  In computing the values presented in
Table~\ref{clusttab}, we have attempted to use the studies with the
largest samples having uniform photometry from the level of the HB to
below the TO. We were able to derive values for the clusters M68
(Walker 1994), M53 and NGC 5053 (Heasley \&
Christian 1991), M92 (Bolte \& Roman 1998), and M15 (Durrell \& Harris
1993).  Although the clusters all have blue HB morphologies, it has
not been necessary to make corrections to find the ``true'' HB level:
either the HB is populated on both sides of the instability strip
(M53, NGC 5053) or there are a number of well-measured RR Lyrae
stars (M15, M68, M92).

From our photometry of M30, we find $V_{TO} = 18.63 \pm 0.05$, and
$\Delta V_{TO}^{HB} = 3.59\pm0.06$.  While M15, M53, M68, M92, and NGC
5053 have $\Delta V_{TO}^{HB}$ values that agree to within the errors
(and also agree with the values derived for clusters of higher
metallicity), M30 has a value about 0.15 mag higher. This is in
disagreement with values given in the extensive tabulation of
Chaboyer, Demarque, \& Sarajedini (1996). The values for M30 and M92
in particular have been put on firmer ground here since consistent
photometry exists from the HB to the TO. Using the more robust V(BTO)
(the apparent magnitude of a point 0.05 mag redder than the turnoff;
Chaboyer et al 1996),
we can compare $\Delta V_{BTO}^{HB}$ values for M30 and M68, which
also has $VI$ data. We find $3.16\pm0.06$ for M30, and $2.96\pm0.02$
for M68.

There are two plausible ways to explain the apparent 0.15 mag excess
in $\Delta V_{TO}^{HB}$ for M30 relative to other clusters.  First,
M30 could be older by $\sim 2$ Gyr. This conclusion would be in
conflict with that inferred by VBS based on the color-difference
method.  (This is perhaps the first case for which the relative age
indicators $\Delta V_{TO}^{HB}$ and the subgiant-branch color extent
give significantly different answers.)  Alternatively, M30 stars could
have a larger initial helium abundance by approximately 0.027. If the
higher helium abundance is restricted to the cluster HB stars, as
would be the case in a deep-mixing scenario, the increase required is
approximately 0.045. (The agreement between the M30 and M68 CMDs
everywhere but on the HB would argue against a difference in the
initial helium abundances in the two clusters, as would beliefs about
Galactic chemical evolution.) The size of the potential helium
enhancement is close to what was inferred earlier from the helium
indicators $\Delta$ and $R$ for M30.

\subsection{Luminosity Functions}

In the following, we will be using a combination of oxygen-enhanced
(BV92) and $\alpha$-element enhanced (VandenBerg 1997) theoretical
LFs to interpret the data. The current state of knowledge
indicates that all of the $\alpha$ elements have enhancements
(Pilachowski, Olszewski, \& Odell 1983; Gratton, Quarta,
\& Ortolani 1986; Sneden et al. 1992). The available evidence also
suggests that the oxygen enhancement remains constant, at least for [Fe/H]
$\lesssim -1.4$ (e.g. Suntzeff 1993, Carney 1996).

On the RGB, stellar evolution is insensitive to the oxygen abundance
because the luminosity evolution is driven almost entirely by the
helium core mass (Refsdal \& Weigert 1970), while the color is primarily
determined by H$^{-}$ opacity. Oxygen has a relatively high ionization
potential, and hence does not contribute electrons to the opacity. The
fainter one goes on the MS, the more insensitive the evolution is to the
oxygen abundance because of the same opacity effect, and because $p-p$
chain reactions are dominant over CNO cycle reactions in influencing
the luminosity.  As a result, the different
distribution of heavy elements causes negligible differences in the
theoretical LFs on the RGB and lower MS (see Figure 17 of Sandquist et
al. 1996).

It is primarily the turnoff region that is affected by changes
in the oxygen abundance, because CNO cycle reactions begin to become important,
and because oxygen ionization regions are close enough to the surface to
influence surface temperatures.  Increased oxygen abundance increases
the envelope opacity, creating redder models. Increased CNO cycle
activity causes a star to adjust to accommodate the increased
luminosity by reducing the temperature and density of the hydrogen
burning regions, which results in a net {\it decrease} in the
luminosity of the turnoff and SGB relative to solar-ratio models. Thus,
the SGB ``jump'' moves in magnitude in the LF. In the CMD, it also
changes slightly in slope, but for metal-poor clusters like M30, this does not
cause a significant change in the shape of the SGB jump in the LF.

An examination of BV92 models indicates that the $V$-band LF is not
very age-sensitive for this range of metallicities. It is most
sensitive on the SGB and then, as found in Sandquist et al. (1996),
only when the SGB is nearly horizontal in the CMD. In
$V$ band for a cluster as metal-poor as M30, the SGB has a relatively
large slope, and so only a large systematic age error will influence the
fit. In light of the {\it Hipparcos} parallax data, this possibility should
be considered, since derived distance moduli indicate brighter TO
magnitudes (and thus, younger ages) for metal-poor clusters.
Figure~\ref{vlfages} shows a comparison of the $V$-band LF with
theoretical LFs for different ages, using an apparent distance modulus
of $(m-M)_V = 14.87$, as derived from one fit to the {\it Hipparcos}
subdwarf sample.

Previous studies of M30's $V$-band LF (Piotto et al. 1987, Bolte 1994,
Bergbusch 1996) uncovered two unusual features in comparisons with
theoretical models: an excess of faint red giants relative to
main-sequence stars, and an excess of subgiant stars.  Our photometry
goes fainter on the MS, allowing us to verify that the normalization of
the theoretical models has not been made in an ``abnormal'' section,
while our wide field allows us to measure the largest sample of red
giants in the cluster to date.

Figure~\ref{vlfcomp} shows a comparison of the studies, with magnitude
shifts according to measured zero-point differences. In large part
there is excellent agreement.  Our LF is significantly below
Bergbusch's at his faint end, most likely due to underestimated
incompleteness corrections in his study. At the bright end of the RGB
($V < 15$), our LF points are also below most of Bergbusch's. However,
we observed a larger number of giants, and our bins are larger, making
our points more significant statistically.

In the following subsections, we discuss the main features in the LFs.

\subsubsection{Red Giant Branch Excesses}

There is a apparently a considerable excess of stars on the RGB for
$15<V<17$ (we will refer to this as the ``lower RGB'') when compared
to the models normalized to the unevolved main sequence. To judge the
reality of the excess, we need to accurately normalize the theoretical
LFs in the horizontal and vertical directions, and choose the
photometry subsample to maximize the statistical significance. The
horizontal normalization can be accomplished by shifting the
theoretical LF in magnitude so that the TO matches that of the
observational LF (Stetson 1991a). In the vertical direction, we have
normalized to the MS in a range of magnitudes where there are large
numbers of measured stars, and where incompleteness is a relatively
small consideration. 

The mass function controls how well the normalized theoretical LF is
fit to the MS portion in the present example. As shown in
Figure~\ref{vlffig}, fits using small values for the power-law mass
function exponent $x$ indicate that the relative numbers of stars on
the RGB and lower MS can be matched by canonical stellar evolution
models. With such a choice though, bins with $18.3 < V < 20.1$ are not
well-modeled. We find that the LF can be modeled from near the faint
limit of our survey to the base of the RGB if we use a higher value
for $x$. This alleviates the depression in the star counts in this
magnitude range seen by Bolte (1994).  However, we are still left with
an excess of RGB stars relative to MS stars in the range $15.1 < V <
16.6$.

The effects of mass segregation have been previously observed within M30 in the
form of a variation of the local mass function exponent $x_{local}$
with radius (RFV, Bolte 1989, Piotto et al. 1990, Sosin
1997). As a result, the best comparison that can be made would be
between theory and a faint sample restricted to the outskirts of the
cluster. The models of Pryor, Smith, \& McClure (1986), as well as
observational studies, indicate that restricting the sample to stars
more than 20 core radii from the center should minimize the effects of
mass segregation on faint end of the LF.

Figure~\ref{vlfrgbms} shows the LF we computed for this purpose.
The presence of the RGB-MS discrepancy in this
case suggests that the problem is not related to the dynamical effects
on the mass function, at least in the outskirts. We can get good
overall agreement with the shape of the LF on the MS, but there is a
relative excess of RGB stars, and the SGB region is not well fit.  The
SGB comparison is insensitive to age and metallicity using this
method of matching the MSTO.  Helium
abundance, however, has a larger effect (Stetson 1991a). Because the
RGB stars in M30 become more populous relative to HB stars (and
presumably MS stars) towards the center, we expect that the
cluster core would show a larger discrepancy.

As with the SGB excess, this effect has only been observed in
metal-poor clusters (in other words, not in the LFs of NGC 288 or M5).
To add stars to the canonical number at a point in the red giant LF,
one must either increase the hydrogen content of the mass being fed
into the hydrogen-burning shell, or reduce the density or temperature
of the burning shell. One possibility for the excess stars on the
lower RGB is that we are seeing the effects of deep mixing, which
brings hydrogen-rich envelope material into the energy generating
shell. If this kind of mixing occurred on the lower RGB, it could
eliminate the RGB bump by erasing the chemical discontinuity left by a
surface convection zone.

Alternately, VandenBerg, Larson, \& DePropris (1998) have examined the
effects of rotation on RGB evolution. They found that core rotation
can expand the outer portions of the stellar core enough to cause a
reduction of the shell temperature. This results in a decrease in the
rate of evolution for RGB stars, and hence leads to an increase in the
number of stars per luminosity bin. This is in the correct direction
to explain the RGB excess. This rotation could be related to
deep mixing scenarios that are required to explain
abundance anomalies in RGB stars (e.g. Shetrone 1996) --- most notably a
decline in the $^{12}$C/$^{13}$C ratio relative to theoretical
predictions, and Na-O and Al-O anti-correlations (as surface material
is mixed into regions where O is being converted to N in the CNO
cycle). [Note that rotation {\it cannot} explain subgiant branch
excesses because the burning region in core-burning stages is too
small to contain a significant amount of angular momentum (VandenBerg
1995). Even if rotation does affect the structure of the star outside
the core, and thereby changes the core temperature, this would not
produce isothermalization that would lead to SGB excesses.]

The rotation and mixing pictures (with the assumption that mixing is
somehow based on internal rotation) receive some support from
observations of rotation in HB stars of some clusters.  There is
definite evidence of stellar rotation in blue HB stars in NGC 288, M3,
and M13 (Peterson, Rood, \& Crocker 1995). M13 has the fastest
rotators, with stars falling into two groups: some with $v \approx 15$
km s$^{-1}$, and some with $v \approx 38$ km s$^{-1}$. M3 has a $v
\sin i$ distribution consistent with $v = 13 \pm 2$ km s$^{-1}$, while
NGC 288's stars are consistent with $v = 9 \pm 2$ km s$^{-1}$. Cohen
\& McCarthy (1997) also found projected rotation rates between 15 and
40 km s$^{-1}$ for five blue HB stars in M92. The presence of stellar
rotation on the HB implies that angular momentum may have been stored
during the RGB phase in a rapidly rotating core, avoiding loss of
angular momentum through the stellar wind. (Such mass loss is needed
to be able to create HB stars of appropriate masses to match observed
cluster HB morphologies.) If this is true, it would be particularly
interesting to compare LFs for M3 and M13 to look for the effects of
rotation, and perhaps even different levels of rotation. Further
stellar rotation measurements for M30, M68, and M92 would also be
helpful in examining rotation as a cause of MS-RGB discrepancy in the
combined LF.

\subsubsection{The RGB Bump}

Figure~\ref{clfhbs} presents the cumulative LF (CLF) for the cluster.
In this graph we have included RGB stars from $1^{\prime}$ to
$6^{\prime}$ from the center of the cluster. The RGB bump is typically
identified from a break in slope in the cumulative LF. At this point,
the shell-burning source begins consuming material of constant, lower
helium content (in other words, the shell reaches what was
formerly the base of the convection zone at its maximum extent ---
Fusi Pecci et al.  1990). Fusi Pecci et al. examined clusters
over a range of metallicities, and found a linear relation between
$\Delta V_{bump}^{HB} = V_{bump} - V_{HB}$ and [Fe/H], as predicted by
theory.  By combining CMDs for three of the most metal-poor clusters
(M15, M92, and NGC 5466), they found $\Delta V_{bump}^{HB} = -0.51 \pm
0.05$. In addition, for NGC 6397, the most metal-poor cluster
for which they were able to find the bump, they found $\Delta
V_{bump}^{HB} = -0.40\pm 0.16$.

As shown in Figure~\ref{clfhbs}, we have examined data for M68 (Walker
1994), a cluster of nearly the same metallicity as M30, in order to
get a better idea of where the bump should be. There is a clear
indication of a slope break for M68: $\Delta V_{bump}^{HB} = -0.46 \pm
0.03$, or $V_{TO} - V = 3.87$, for M68. That result shows that the
continuation of the Fusi Pecci et al. relation to lower metallicity
appears correct. We have chosen to shift M30 and M68 so that their
MSTOs align because of the evidence that M30's HB may be anomalously
bright (see \S~\ref{dist}). The comparison reveals that there may be a
feature at the same position as in M68, although 
we do not see significant signs of slope change in the CLF at the
position of the feature. 

\subsubsection{Subgiant Branch Excesses}

We find that a few bins on the SGB ($V \approx 18.17$)
show an excess of stars relative to the theoretical predictions for
the best fitting models, confirming the result of Bolte
(1994). In Figure~\ref{vlfsgb}, we plot the LF with a radius cut closer to the
cluster center so as to get better statistics on the SGB. As
Figure~\ref{vkept} shows, there is little scatter in the vicinity of
the SGB in the CMD that would tend to wash out or contribute to the
observed excess at $V \approx 18.2$. The excess is based on a single
point having a significance of $2.7 \sigma$, where the error in almost
entirely due to Poisson statistics. Bolte (1994) states the
significance of the bump as $4.8 \sigma$, and it appears to occupy two
LF bins in his Figure~7.  The significance of his result is probably
smaller than that because of the difficulty in determining the
position of the ``jump'' ($\approx 0.8$ mag brighter than the MSTO in
$V$) in his LF.  

An examination of the $I$-band LF in Figure~\ref{ilf} shows the
presence of a deviation at the same position in the
CMD. This is important because the slope of the SGB is
steeper in an $(I,V-I)$ CMD than in a $(V,V-I)$ CMD. As a result, the
bump can no longer be ascribed to a feature caused by the
exact slope: it must be the result of an increase in the number of
stars congregating near a point on the cluster's fiducial line.
At the analogous position in the $I$-band LF, there
are two bins with excesses of $1.7 \sigma$ and $2.8 \sigma$ compared
to theory, for a combined significance of $3.3 \sigma$.  The
appearance of the subgiant branch excess in both the $V$- and $I$-band
LFs indicates that the cause must be due to an excess of stars (rather
than being caused by the exact slope of the SGB --- thus eliminating
the exact metallicity, helium content, and oxygen abundance
as causes).

So, the SGB excess has marginal significance in our LFs. In order to
more definitively determine the reality of the feature, photometry
reaching into the center of this cluster will be needed. The
observation of this feature in the combined LF of M68, M92, and NGC
6397 (Stetson 1991a) lends more credence to the phenomenon, but more
investigation is necessary.

Sandquist et al. (1996) found that there was no evidence
for an SGB excess in the LF of the more metal rich cluster M5 (which
has a good $I$-band LF for easy comparison with Figure~\ref{ilf}).
Bergbusch (1993) saw no evidence of an excess in his $V$-band LF of
NGC 288. These pieces of evidence seem to indicate that any
cause must only be effective at low
metallicities.  There is, however, a general lack of useful LF data
covering the SGB for globular clusters with metallicities between M5
and M30, or more metal rich than M5.

If the feature is real, there are at least two potential means of
creating such an excess: a fluctuation in the initial mass function,
and an unknown physical process isothermalizing the stellar core of
turnoff mass stars. The excess in Stetson's M68-M92-NGC 6397 LF makes
mass function fluctuations less likely.  A star can be forced to pause
on the SGB, but still burning hydrogen in its core, if isothermality is imposed
on a large portion of the core (Faulkner \& Swenson 1993). If such a
process occurred in a large enough fraction of the stars in a cluster,
a SGB excess could be created in the LF.  A way to create such an
excess is to invoke a process that increases the efficiency of energy
transfer over a large portion of the core. For this to happen, the
mean free path of the transporting particle must be large. No such
particle has been identified to date.

\section{Conclusions}\label{conc}

1. Determinations of the reddening for M30 disagree at a $\pm 0.03$
mag level and cases can be made for values ranging from 0.03 to 0.07
in E$(V-I$). This uncertainty is the main factor preventing a more
accurate determination of the distance modulus. By fitting subdwarfs
with {\it Hipparcos} parallax data to the $VI$ fiducial line, we find
satisfactory fits for $(m-M)_{V}$; E($V-I$) pairs ranging from 14.87;
0.06 to 14.65; 0.02 with the statistical errors of around 0.12 mag
(all for the case [Fe/H] $= -1.91$). When shifted to a common
reddening, we find our distance modulus is consistent with
ground-based estimates, and at least 0.1 mag smaller than other
Hipparcos-based estimates.

2. M30 has a larger $R$ value [$N_{HB}/N_{RGB}$] than any of the other
metal-poor clusters for which this quantity has been measured.
This quantity is usually used as a helium indicator and our measured
$R$ value
suggests a helium abundance $\sim 0.03-0.04$ larger than
the mean of the other metal-poor clusters.
M30's value for the helium indicator $\Delta$ is also relatively high
although for the case of E$(V-I) <0.02$, it is consistent with 
the other metal-poor clusters. If there is a helium abundance
enhancement in M30, it is probably not an initial abundance difference
since Galactic chemical evolution and the similarity of the M30 and M68
fiducial lines (see next point) argue against it.
 
3. The $\Delta V_{TO}^{HB}$ value for M30 is demonstrably 
large relative to clusters of similar metallicity. The M30 
fiducial line (except for the HB) overlies that of 
M68 (see Figure~\ref{m30vsm68}) and
M92 (VBS) very closely, indicating that M30 probably has the same age as these
two clusters. We suggest that the HB luminosity in M30 is high
due to a larger-than-average Y for the M30 HB stars.

4. The LFs of the cluster show definite evidence for an excess of RGB
stars relative to MS stars, and marginally significant ($\approx 3
\sigma$) evidence for an excess of SGB stars, as compared with theory.
The SGB feature has slightly higher significance in the $I$ band.  The
possibility remains that these anomalies are only present in
low-metallicity clusters.

Stellar rotation is a possible explanation for the excess number
of RGB stars relative to MS stars. Alternatively, the excess giants 
could be a signpost for mixing events on the lower RGB in which
fresh hydrogen is mixing into the energy generation region. This could
also be identified as the source of the envelope Y-enrichment
we infer from the HB and brighter RGB stars.

5. We do not find an obvious RGB bump in M30, in spite
of the size of our RGB sample. Using the cumulative LF, we have
detected the bump in the metal-poor cluster M68 with a
$\Delta V_{bump}^{HB}$ value that agrees with the
linear trend with [Fe/H] found by Fusi Pecci et al. (1990). 

It is possible that points 2 -- 5 are all related to the deep-mixing
events inferred for some globulars based on the surface abundances of
elements that participate in the energy generation cycles.  The
hypothesis that we are seeing the effects of the mixing of
hydrogen-rich material into the energy-generation regions and
helium-rich material out into the stellar envelope can qualitatively
explain all of these (2 through 5) observations. If this hypothesis is
correct then we predict that detailed abundance studies of the bright
giants in M30 should show the characteristic patterns of deep mixing
-- low oxygen and carbon abundances accompanied by high nitrogen,
aluminum, and sodium.

\acknowledgments

We would especially like to thank D. VandenBerg for providing us with
theoretical $\alpha$-enhanced isochrones and luminosity functions
prior to publication and P. Stetson for the use of his excellent
software. It is a pleasure to thank P. Guhathakurta, Z.
Webster, and R. Rood for useful conversations.  This research has
made use of the SIMBAD database, operated at CDS, Strasbourg, France.
M.B. is happy to acknowledge support from NSF grant AST 94-20204.

Electronic copies of the listing of the photometry are available on
request to the first author.

\newpage

\appendix

\section{Data Reduction}

\subsection{Primary Standard Calibration Fields}

\subsubsection{Aperture Photometry}

Aperture photometry was performed using the program DAOPHOT II
(Stetson 1987). Using these data, growth curves were constructed for
each frame using DAOGROW (Stetson 1990) in order to extrapolate from
the flux measurements over a circular area of finite radius to the
total flux observable for the star.  The aperture magnitudes and the
known standard system magnitudes of Landolt (1992) were then used to
derive coefficients for the transformation equations:
\[ v = V + a_{0} + a_{1} \cdot (X - 1.25) + a_{2} \cdot (V - I)+ a_{3}
\cdot (V - I)^{2} + a_{4} \cdot (V - I)^{3}  \]
\[ i = I + b_{0} + b_{1} \cdot (X - 1.25) + b_{2} \cdot (V - I) ,\]
where $v$ and $i$ are observed aperture photometry magnitudes, $V$ and
$I$ are the standard system magnitudes, and $X$ is the airmass.  The
primary standard stars covered a color range $-0.35
< (V - I) < 1.67$, completely encompassing the color range of the
cluster sample. The coefficients for the transformation equations are
given in Table~\ref{coeftab}.  The residuals for the sample of 25
stars are shown in Figure~\ref{prime}, and the average residuals are
given in Table~\ref{restab}.  (In this and all subsequent comparisons, the
residuals are calculated in the sense of ours -- theirs.)

\subsection{Secondary Standard Calibration}

We chose 118 relatively bright and isolated stars in the M30 field
observed with the 1.5~m telescope during the photometric night to be
``secondary standards''.  The stars selected were required to be
unsaturated, brighter than the turnoff, in relatively uncrowded
regions of the images, and close to the apparent fiducial line of the
cluster (since this acts as an additional check on the accuracy of the
photometry).  Once the list was finalized, all other stars were
subtracted from the frames and aperture photometry was obtained. The
colors for these standards cover the range $12.82 < V < 16.54$
and $-0.069 < (V-I) < 1.280$

\subsection{Object Frames}\label{3.3.}

\subsubsection{Profile Fitting Photometry}\label{psf}

Both the CTIO 4~m and 1.5~m data for M30 were reduced using the
standard suite of programs developed by Peter Stetson
(DAOPHOT/ALLSTAR; Stetson 1987, 1989), and following the
procedures in Sandquist et al. (1996).

\subsubsection{Calibration}\label{cal}

We used the secondary standards established with the 1.5~m
observations on the single photometric night to determine the
coefficients in the transformation equations for all of the 1.5~m profile
fitting photometry:
\[ v = V + a_{0,k} + a_{1} \cdot (V - I) \]
\[ i = I + b_{0,k} + b_{1} \cdot (V - I) ,\]
where $v$ and $i$ are the instrumental magnitudes from the profile
fitting, $V$ and $I$ are the standard values from the aperture
photometry, and $k$ is an index referring to individual frames.
The coefficients of the color terms are given in
Table~\ref{coeftab}, the average residuals and standard deviation of
the residuals for the comparison of the profile
fitting and aperture photometry are given in Table~\ref{restab}, and individual
star residuals are shown in Figure~\ref{aptopsf}.

In the next step, we chose to calibrate the 4~m profile fitting
photometry to the 1.5~m profile fitting photometry rather than the
aperture photometry of the secondary standards. This was done
primarily to ensure that all of our profile fitting was on the same
system over as large a range of magnitudes as possible. The M30 frames
taken at the 1.5~m telescope on the one photometric night did not go
particularly deep, while the 4 m photometry had few unsaturated observations of
the brighter stars in the cluster. The data taken on the
non-photometric nights at the 1.5 m telescope did, however, cover a
range of magnitudes similar to that of the 4 m data. 

We selected a sample of 248 stars found in both fields at least 300
pixels away from the cluster center. These stars were used
to determine the transformation coefficients for the equations:
\[ v = V + a_{0,k} + a_{1} \cdot (V - I) + a_{2} \cdot (V - I)^{2}\]
\[ i = I + b_{0,k} + b_{1} \cdot (V - I) + b_{2} \cdot (V - I)^{2},\]
where $v$ and $i$ are the instrumental magnitudes from the 4~m
observations, and $V$ and $I$ are the standard values from the 1.5~m
observations.  The coefficients of the color terms are given in
Table~\ref{coeftab}, while the residuals of the comparison of the
photometry for the 4~m and 1.5~m measurements of the secondary
standards are shown in Figure~\ref{secres}.

For the final calibration, we used the transformation equations for
the 1.5~m and 4~m profile-fitting data. All
of the profile-fitting photometry from both telescopes was combined
with weights equal to the inverse square of the internal measurement
errors in order to determine our standard-system magnitude and
color values.

Because of the large sky coverage of the CTIO frames, most other
surveys of M30 overlap the program area at least partially.
Table~\ref{restab} provides a summary of the zero-point offsets for
comparisons with these studies. We would particularly like to point
out that there is considerable difference among them,
highlighting the importance of the calibration. The fields used by
Bolte (1987), RFV, and Samus et al. (1995) are completely included on
all frames. A comparison with the photometry of Bolte is given in
Figure~\ref{boltecomp}. In Figure~\ref{alccomp}, we show the
comparison with the study of M30 by Samus et al. (1995). We do this
partly because it involves the same filter bands, and partly because
the residuals are the lowest on average (although the scatter in
star-to-star residuals is large). As a note, comparisons with the most
recent study (Bergbusch 1996) show no signs of color trends in the
residuals except within about a magnitude of the tip of the giant
branch.

\newpage

\figcaption{ The M30 color-magnitude
diagram for a) all measured stars, and b) the sample restricted to stars
having $110^{\prime\prime} <r< 10^{\prime}$. \label{m30cmd}}

\figcaption{ Magnitude bias versus $V$ magnitude. \label{delta}}

\figcaption{ External magnitude errors versus $V$
magnitude. \label{esigma}}

\figcaption{ Completeness fraction versus $V$
magnitude. \label{fsfig}}

\figcaption{Subdwarf fits to the $VI$ main sequence fiducial
 for two different reddenings, assuming [Fe/H]$_{M30}=-1.91$ (Carretta
\& Gratton 1996).  The M30
fiducial line is plotted as open boxes. The solid boxes with error
bars are local subdwarfs and giants. Only those subdwarfs with
$M_V>4.5$ and [Fe/H]$< -1.2$ were used in the main-sequence fitting.
The isochrones (plotted as solid lines) are preliminary
$\alpha$-enhanced versions (VandenBerg 1997) with [Fe/H] $ = -2.01$
and ages 10, 12, 14, and 16 Gyr (from top to bottom). The two
reddening values correspond to E$(B-V)=0.016$ and 0.048.
\label{subdvred}}

\figcaption{The M30 cumulative radial distributions for RGB stars ({\it
solid line}) HB stars ({\it dotted line}), and AGB stars ({\it dashed
line}), using wide-field data from this paper, and HST data for the
core (YGSB). \label{crd}}

\figcaption{ A comparison of the $(V,V-I)$ fiducial lines for
M30 ({\it thick line}) with our computed fiducials from M68 ({\it thin
line}; Walker 1994 data). The M68 fiducial has been shifted 0.45 mag brighter
in $V$ and 0.03 mag bluer in color. \label{m30vsm68}}

\figcaption{The $V$-band luminosity function for M30
compared with theoretical $\alpha$-enhanced LFs for [Fe/H] $=
-2.01$ and a distance modulus
$(m-M)_{V}=14.87$ for ages (from left to right 10, 12, 14, and 16 Gyr.
The theoretical luminosity functions have been
normalized to the range $20.4 < V < 20.9$. \label{vlfages}}

\figcaption{The $V$-band luminosity function for 14772 stars in M30 as
derived here ($\bullet$ with error bars), compared with LFs from
Bergbusch (1996; $\triangle$) and Bolte (1994; $\Box$), as well as a
theoretical $\alpha$-enhanced LFs for [Fe/H] $= -2.01$ 
and $x=2.0$ with age 12 Gyr and $(m-M)_{V}=14.70$ ({\it solid line})
and age 10 Gyr and $(m-M)_{V}=14.87$ ({\it dotted line}) for
comparison purposes. All of the luminosity functions have been
normalized to the range $18.5 < V < 19$. \label{vlfcomp}}

\figcaption{The $V$-band luminosity function in M30,
with theoretical $\alpha$-enhanced LFs for [Fe/H] $= -2.01$ and age 12
Gyr. The curves are for mass function exponents $x = 1.5$ ({\it solid
line}), $x = 2$ ({\it dotted line}), and $x = 2.5$ ({\it dashed line}).
The theoretical LFs have been shifted in magnitude using an apparent
distance modulus $(m-M)_{V} = 14.70$. \label{vlffig}}

\figcaption{The $V$-band luminosity function for M30, restricted to $r >
4^{\prime} \approx 30 r_{c}$
with theoretical $\alpha$-enhanced LFs for [Fe/H] $= -2.01$, and $x=2$.
The four curves are for ages 10, 12, 14, and 16 Gyr, and they have
been shifted so that the theoretical and observational TO magnitudes
match, and then normalized to the
observed LF in a 0.5 mag bin 2 magnitudes below the turnoff. The
implied distance moduli from the isochrone TOs are 14.98, 14.77,
14.61, and 14.47 respectively.\label{vlfrgbms}}

\figcaption{The cumulative luminosity functions for M30 and M68 (Walker
1994), shifted so that the main sequence turnoffs have the same
magnitude. The dotted lines indicate linear fits to the functions
above and below the slope break, which is an indication of the
position of the RGB bump. \label{clfhbs}}

\figcaption{The $V$-band luminosity function for M30
including stars down to a radius 150 pixels ($66^{\prime\prime}$) from
the cluster center. The theoretical
$\alpha$-enhanced LFs have [Fe/H] $= -2.01$ and $x=2.5$, but have the
same combinations of age and distance modulus as Figure~\ref{vlfcomp}.
The luminosity function has been normalized to the range $19.4 < V < 19.9$.
\label{vlfsgb}}

\figcaption{ The M30 color-magnitude
diagram for all the stars in the $VI$ sample used in computing the
$V$-band luminosity function in Figure~\ref{vlfsgb}. \label{vkept}}

\figcaption{The $I$-band luminosity function
for 14507 stars, with [Fe/H] $=
-2.01$, $x = 2.5$ theoretical $\alpha$-enhanced LFs for age 12 Gyr and
$(m-M)_{I} = 14.67$ ({\it solid line}), and age 10 Gyr and $(m-M)_{I} =
14.81$ ({\it dotted line}). \label{ilf}}

\figcaption{Final residuals for the comparison of standard and
measured values for Landolt (1992) primary standard stars observed at
the CTIO 1.5~m telescope. \label{prime}}

\figcaption{Residuals for the comparison of 118 M30 secondary
standard stars measured using profile fitting and aperture photometry
in the CTIO 1.5~m frames.  The residuals are in
the sense (profile fitting -- aperture).\label{aptopsf}}

\figcaption{Final residuals for the comparison of 248 M30 
stars used in the calibration of the CTIO 4~m frames. The residuals are in
the sense (4~m -- 1.5~m).\label{secres}}

\figcaption{Residuals for the comparison between the M30 CCD photometry of
Bolte (1987) and our dataset. The data for Bolte's short exposure
frames are in the plots on the left, and his long exposure data is
used on the right. The residuals are in the sense (ours -- Bolte's).
\label{boltecomp}}

\figcaption{Residuals for the comparison between the M30 CCD photometry of
Samus et al. (1995) and our dataset. The residuals are in the sense
(ours -- Samus').\label{alccomp}}

\newpage

\begin{deluxetable}{cccccc}
\tablecolumns{6}
\tablewidth{0pc}
\tablenum{1}
\tablecaption{M30 [$V, (V-I)$] Fiducial Points}
\tablehead{\colhead{$V$} & \colhead{$V-I$} & \colhead{N} &
\colhead{$V$} & \colhead{$V-I$} & \colhead{N}}
\startdata
\multicolumn{3}{c}{MS-SGB-RGB} & 17.275 & 0.820 & 29 \nl
23.000 & 1.293 &  303 & 17.216 & 0.824 &   28 \nl
22.800 & 1.293 &  399 & 17.077 & 0.833 &   20 \nl
22.600 & 1.209 &  556 & 16.925 & 0.840 &   23 \nl
22.400 & 1.147 &  771 & 16.781 & 0.847 &   14 \nl
22.200 & 1.115 &  918 & 16.618 & 0.854 &   38 \nl
22.000 & 1.052 & 1048 & 16.391 & 0.863 &   35 \nl
21.925 & 1.035 &  827 & 16.140 & 0.880 &   40 \nl
21.775 & 1.002 &  811 & 15.868 & 0.901 &   25 \nl
21.625 & 0.967 &  850 & 15.621 & 0.913 &   37 \nl
21.475 & 0.928 &  896 & 15.367 & 0.927 &   18 \nl
21.325 & 0.890 &  934 & 15.127 & 0.953 &   14 \nl
21.175 & 0.869 &  907 & 14.877 & 0.969 &   14 \nl
21.025 & 0.837 &  910 & 14.662 & 0.990 &   11 \nl
20.875 & 0.810 &  965 & 14.395 & 1.013 &  6 \nl
20.725 & 0.779 &  908 & 14.117 & 1.052 &  8 \nl
20.575 & 0.748 &  876 & 13.924 & 1.068 &  6 \nl
20.425 & 0.719 &  856 & 13.604 & 1.115 &  3 \nl
20.275 & 0.699 &  843 & 13.319 & 1.143 &  3 \nl
20.125 & 0.675 &  740 & 13.125 & 1.178 &  7 \nl
19.975 & 0.672 &  765 & 12.830 & 1.223 &  3 \nl
19.825 & 0.646 &  694 & 12.634 & 1.260 &  3 \nl
19.675 & 0.634 &  620 & 12.384 & 1.329 &  1 \nl
19.525 & 0.622 &  566 &	12.016 & 1.481 &  3 \nl
19.375 & 0.605 &  519 & \multicolumn{3}{c}{HB} \nl
19.225 & 0.588 &  444 & 15.821 & -0.016 & 6 \nl
19.000 & 0.576 &  527 & 15.584 & 0.022 & 6  \nl
18.800 & 0.568 &  466 & 15.386 & 0.084 & 4 \nl
18.600 & 0.558 &  371 & 15.302 & 0.124 & 3 \nl
18.400 & 0.575 &  305 & 15.187 & 0.169 & 2 \nl
18.240 & 0.595 &  272 & 15.083 & 0.306 & 5  \nl
18.094 & 0.645 &   95 & 14.900 & 0.701 & 6  \nl
17.968 & 0.695 &   53 & \multicolumn{3}{c}{AGB} \nl
17.884 & 0.745 &   56 &14.184 & 0.970 & 4 \nl
17.800 & 0.764 &  101 &14.354 & 0.924 & 1  \nl
17.575 & 0.793 &   57 &14.483 & 0.914 & 1  \nl
17.425 & 0.809 &   44 &14.564 & 0.894 & 1  \nl
\enddata
\label{fidtab}
\end{deluxetable}

\begin{deluxetable}{cccccccccc}
\tablecolumns{10}
\tablewidth{0pc}
\tablenum{2}
\tablecaption{$V$-Band Luminosity Function for M30}
\tablehead{\colhead{$V$} & \colhead{$N$} & \colhead{$\sigma(N)$} &
\colhead{$N_{obs}$} & \colhead{$\overline{f}$} & \colhead{$V$} &
\colhead{$N$} & \colhead{$\sigma(N)$} & \colhead{$N_{obs}$} &
\colhead{$\overline{f}$}}
\startdata
12.543 & 0.766 & 0.212 & 13 & 1.000 & 18.855 & 191.524 & 14.548 & 184 & 0.960 \nl
13.522 & 1.028 & 0.265 & 15 & 1.000 & 19.005 & 240.516 & 16.396 & 229 & 0.952 \nl
14.275 & 1.957 & 0.449 & 19 & 1.000 & 19.155 & 276.068 & 17.539 & 261 & 0.945 \nl
14.727 & 2.885 & 0.771 & 14 & 1.001 & 19.305 & 285.390 & 17.865 & 267 & 0.934 \nl
15.027 & 4.539 & 0.968 & 22 & 1.001 & 19.455 & 332.817 & 19.475 & 306 & 0.918 \nl
15.328 & 6.614 & 1.169 & 32 & 0.998 & 19.605 & 361.638 & 20.207 & 331 & 0.915 \nl
15.554 & 7.433 & 1.752 & 18 & 1.001 & 19.756 & 386.727 & 20.903 & 351 & 0.907 \nl
15.704 & 5.785 & 1.546 & 14 & 1.000 & 19.906 & 451.675 & 22.748 & 403 & 0.892 \nl
15.854 & 8.263 & 1.848 & 20 & 1.000 & 20.056 & 475.933 & 23.433 & 419 & 0.880 \nl
16.004 & 9.912 & 2.023 & 24 & 1.000 & 20.206 & 556.464 & 25.362 & 488 & 0.877 \nl
16.154 & 8.259 & 1.847 & 20 & 1.000 & 20.356 & 574.396 & 25.920 & 498 & 0.867 \nl
16.304 & 7.433 & 1.752 & 18 & 1.001 & 20.506 & 632.285 & 27.495 & 537 & 0.850 \nl
16.455 & 11.985 & 2.226 & 29 & 1.000 & 20.656 & 664.107 & 28.509 & 555 & 0.837 \nl
16.605 & 11.161 & 2.148 & 27 & 1.000 & 20.805 & 698.896 & 29.602 & 575 & 0.824 \nl
16.755 & 12.813 & 2.302 & 31 & 1.000 & 20.955 & 733.417 & 30.831 & 592 & 0.809 \nl
16.905 & 16.977 & 4.117 & 17 & 1.001 & 21.105 & 781.575 & 32.377 & 622 & 0.798 \nl
17.055 & 12.982 & 3.601 & 13 & 1.001 & 21.254 & 751.962 & 32.336 & 588 & 0.784 \nl
17.205 & 17.971 & 4.236 & 18 & 1.001 & 21.404 & 805.608 & 33.886 & 620 & 0.774 \nl
17.355 & 25.019 & 5.004 & 25 & 0.998 & 21.553 & 779.031 & 34.422 & 580 & 0.749 \nl
17.505 & 16.997 & 4.122 & 17 & 0.999 & 21.702 & 903.370 & 38.618 & 648 & 0.722 \nl
17.655 & 34.968 & 5.912 & 35 & 1.000 & 21.851 & 922.898 & 40.324 & 643 & 0.702 \nl
17.806 & 50.947 & 7.140 & 51 & 1.000 & 22.000 & 986.270 & 44.897 & 643 & 0.658 \nl
17.956 & 63.013 & 7.972 & 63 & 1.000 & 22.148 & 998.885 & 46.770 & 622 & 0.631 \nl
18.105 & 102.573 & 10.276 & 102 & 0.997 & 22.296 & 1036.098 & 48.374 & 601 & 0.584\nl 
18.255 & 117.435 & 11.112 & 116 & 0.990 & 22.445 & 1134.865 & 55.105 & 567 & 0.505 \nl
18.405 & 123.294 & 11.517 & 121 & 0.983 & 22.593 & 1350.132 & 71.038 & 501 & 0.378 \nl
18.555 & 159.431 & 13.317 & 155 & 0.973 & 22.740 & 3956.208 & 404.959 & 500 & 0.129 \nl
18.705 & 191.698 & 14.697 & 185 & 0.964 & & & & & \nl
\enddata
\label{vlftab}
\end{deluxetable}

\begin{deluxetable}{cccccccccc}
\tablecolumns{10}
\tablewidth{0pc}
\tablenum{3}
\tablecaption{$I$-Band Luminosity Function for M30}
\tablehead{\colhead{$I$} & \colhead{$N$} & \colhead{$\sigma(N)$} &
\colhead{$N_{obs}$} & \colhead{$\overline{f}$} & \colhead{$I$} &
\colhead{$N$} & \colhead{$\sigma(N)$} & \colhead{$N_{obs}$} &
\colhead{$\overline{f}$}}
\startdata
11.149 & 0.590 & 0.170 & 12 & 1.000 &  17.467 & 76.966 & 8.715 & 78 & 1.000 \nl
12.225 & 0.850 & 0.236 & 13 & 1.000 &  17.619 & 111.136 & 10.508 &  112 & 0.999  \nl
12.993 & 1.281 & 0.355 & 13 & 1.000 &  17.770 & 104.511 & 10.283 &  104 & 0.992  \nl
13.452 & 2.373 & 0.685 & 12 & 1.000 &  17.921 & 131.605 & 11.657 &  130 & 0.985  \nl
13.756 & 2.779 & 0.743 & 14 & 1.000 &  18.071 & 188.224 & 14.172 &  183 & 0.972  \nl
14.060 & 3.586 & 0.845 & 18 & 1.000 &  18.221 & 185.096 & 14.125 &  178 & 0.964  \nl
14.287 & 4.359 & 1.315 & 11 & 1.001 &  18.370 & 253.828 & 16.588 &  243 & 0.959  \nl
14.439 & 6.772 & 1.643 & 17 & 0.997 &  18.520 & 290.213 & 17.909 &  273 & 0.943  \nl
14.591 & 7.986 & 1.786 & 20 & 1.001 &  18.670 & 302.430 & 18.308 &  282 & 0.936  \nl
14.742 & 5.598 & 1.496 & 14 & 1.000 &  18.819 & 345.452 & 19.861 &  313 & 0.911  \nl
14.893 & 5.998 & 1.549 & 15 & 1.000 &  18.968 & 421.574 & 21.974 &  380 & 0.905  \nl
15.044 & 6.780 & 1.645 & 17 & 1.000 &  19.117 & 465.390 & 23.205 &  414 & 0.894  \nl
15.195 & 9.954 & 1.991 & 25 & 1.000 &  19.266 & 506.015 & 24.377 &  441 & 0.881  \nl
15.347 & 5.985 & 1.546 & 15 & 1.000 &  19.415 & 613.201 & 27.068 &  526 & 0.860  \nl
15.498 & 9.583 & 1.956 & 24 & 1.000 &  19.564 & 693.000 & 29.054 &  583 & 0.853  \nl
15.649 & 8.810 & 1.879 & 22 & 1.000 &  19.712 & 777.236 & 31.180 &  637 & 0.832  \nl
15.799 & 8.811 & 1.879 & 22 & 1.000 &  19.859 & 820.648 & 32.433 &  657 & 0.819  \nl
15.950 & 12.407 & 2.229 & 31 & 1.000 &  20.006 & 833.922 & 33.018 &  656 & 0.807  \nl
16.101 & 11.995 & 2.190 & 30 & 1.001 &  20.152 & 961.583 & 36.406 &  730 & 0.779  \nl
16.252 & 14.400 & 2.400 & 36 & 1.000 &  20.298 & 1024.388 & 38.171 &  758 & 0.762  \nl
16.403 & 14.773 & 2.429 & 37 & 1.000 &  20.443 & 1042.008 & 39.215 &  755 & 0.748  \nl
16.554 & 21.175 & 2.909 & 53 & 0.999 &  20.587 & 1104.607 & 42.537 &  743 & 0.708  \nl
16.705 & 21.109 & 2.900 & 53 & 1.000 &  20.730 & 1145.276 & 43.849 &  759 & 0.695  \nl
16.857 & 25.699 & 5.040 & 26 & 1.001 &  20.873 & 1314.175 & 49.073 &  833 & 0.666  \nl
17.008 & 26.650 & 5.129 & 27 & 1.000 &  21.013 & 1385.530 & 55.311 &  777 & 0.610  \nl
17.161 & 41.972 & 6.400 & 43 & 1.000 &  21.155 & 1525.717 & 70.014 &  738 & 0.501  \nl
17.315 & 46.912 & 6.771 & 48 & 1.000 & & & & & \nl
\enddata
\label{ilftab}
\end{deluxetable}

\begin{deluxetable}{lcccccccl}
\tablecolumns{9}
\tablewidth{0pc}
\tablenum{4}
\tablecaption{Metal-Poor Field Subdwarfs and Subgiants}
\tablehead{\colhead{HIP No.} & \colhead{$V$} &
\colhead{$(V-I)$} & \colhead{$\pi$} & \colhead{$\sigma_{\pi}/\pi$} &
\colhead{[Fe/H]} & \colhead{$M_{V}$ \tablenotemark{a}} &
\colhead{$(V-I)_{-1.91}$} & \colhead{Name}}
\startdata
\multicolumn{9}{c}{Subdwarfs} \nl
14594 & 8.04   & 0.66 & 0.02585 & 0.044 & $-1.88$ & $5.08\pm0.10$ & 0.66 & HD19445 \nl 
18915 &   8.51 & 1.01 & 0.05414 & 0.020 & $-1.69$ & $7.18\pm0.04$ & 0.99 & HD25329 \nl 
24316 &	  9.43 & 0.65 & 0.01455 & 0.069 & $-1.44$ & $5.19\pm0.15$ & 0.61 & HD34328 \nl 
38541 &	  8.27 & 0.77 & 0.03529 & 0.029 & $-1.60$ & $6.00\pm0.06$ & 0.75 & HD64090 \nl 
40778 &	  9.73 & 0.60 & 0.01036 & 0.142 & $-1.49$ & $4.64\pm0.31$ & 0.57 & BD+54 1216 \nl 
53070 &	  8.22 & 0.63 & 0.01923 & 0.059 & $-1.38$ & $4.61\pm0.13$ & 0.59 & HD94028 \nl 
57939 &	  6.44 & 0.89 & 0.10921 & 0.007 & $-1.22$ & $6.63\pm0.02$ & 0.84 & HD103095 \nl 
60632 &	  9.66 & 0.63 & 0.01095 & 0.118 & $-1.55$ & $4.73\pm0.26$ & 0.60 & HD108177 \nl 
74234 &	  9.46 & 1.01 & 0.03368 & 0.050 & $-1.57$ & $7.08\pm0.11$ & 0.99 & HD134440 \nl 
74235 &	  9.08 & 0.92 & 0.03414 & 0.040 & $-1.57$ & $6.74\pm0.09$ & 0.90 & HD134439 \nl 
98020 &	  8.83 & 0.75 & 0.02532 & 0.046 & $-1.37$ & $5.77\pm0.10$ & 0.70 & HD188510 \nl 
100568 &  8.66 & 0.67 & 0.02288 & 0.054 & $-1.00$ & $5.43\pm0.12$ & 0.60 & HD193901 \nl 
100792 &  8.35 & 0.63 & 0.01794 & 0.069 & $-1.02$ & $4.58\pm0.15$ & 0.55 & HD194598 \nl 
104659 &  7.37 & 0.66 & 0.02826 & 0.036 & $-0.94$ & $4.59\pm0.08$ & 0.56 & HD201891 \nl 
\multicolumn{9}{c}{Subgiants} \nl
3026 &	  9.25 & 0.64 & 0.00957 & 0.144 & $-1.17$ & $3.97\pm0.31$ & 0.54 & HD3567 \nl 
33221 &	  9.07 & 0.63 & 0.00911 & 0.111 & $-1.33$ & $3.77\pm0.24$ & 0.53 & CPD-33 3337 \nl 
48152 &	  8.33 & 0.55 & 0.01244 & 0.085 & $-2.07$ & $3.69\pm0.18$ & 0.55 & HD84937 \nl 
55790 &	  9.07 & 0.63 & 0.01099 & 0.135 & $-1.56$ & $4.03\pm0.29$ & 0.55 & HD99383 \nl 
68464 &	  8.73 & 0.64 & 0.00977 & 0.135 & $-1.75$ & $3.53\pm0.29$ & 0.60 & HD122196 \nl 
76976 &	  7.22 & 0.69 & 0.01744 & 0.056 & $-2.38$ & $3.32\pm0.13$ & 0.77 & HD140283 \nl 
\enddata
\tablenotetext{a}{Lutz-Kelker corrections calculated using $\Delta M_{LK} =
-7.60(\sigma_{\pi} / \pi)^{2} - 47.23(\sigma_{\pi}/\pi)^{4}$}
\label{subdtab}
\end{deluxetable}

\begin{deluxetable}{cccccc}
\tablecolumns{6}
\tablewidth{0pc}
\tablenum{5}
\tablecaption{Measured Distance Moduli}
\tablehead{\colhead{[Fe/H]} & \colhead{E$(V-I)$} &
\multicolumn{4}{c}{Subdwarf Sample} \\[.2ex]
\cline{3-6}\\ \colhead{} & \colhead{} &
\colhead{All} & \colhead{$M_{V}>4.25$} & \colhead{$M_{V}>5$} &
\colhead{[Fe/H]$<-1.3$}}
\startdata
$-1.91$ & 0.06 & $14.93\pm0.12$ & {\bf $14.87\pm0.12$} & $14.86\pm0.11$ & $14.92\pm0.11$ \nl
$-1.91$ & 0.02 & $14.70\pm0.12$ & {\bf $14.65\pm0.12$} & $14.64\pm0.11$ & $14.76\pm0.12$ \nl
$-2.13$ & 0.06 & $14.93\pm0.12$ & $14.91\pm0.12$ & $14.87\pm0.11$ & $14.93\pm0.11$ \nl
$-2.13$ & 0.02 & $14.69\pm0.13$ & $14.68\pm0.13$ & $14.66\pm0.12$ & $14.71\pm0.13$ \nl
\enddata
\label{dmtab}
\end{deluxetable}

\begin{deluxetable}{lccccc}
\tablecolumns{6}
\tablewidth{0pc}
\tablenum{6}
\tablecaption{Characteristics of Metal-Poor Globular Clusters}
\tablehead{\colhead{ID} & \colhead{[Fe/H]$_{ZW}$} &
\multicolumn{2}{c}{$Y$ Indicators} & \colhead{$\Delta V_{TO}^{HB}$} &
\colhead{$R_{HB}$} \\[.2ex] \cline{3-4} \\
\colhead{} & \colhead{} & \colhead{$R$} & \colhead{$\Delta_{B-V}$} &
\colhead{} & \colhead{}}
\startdata
NGC 104 (47 Tuc) & $-0.71$ &
$1.21\pm0.13$ & $5.32\pm0.08$ & $3.59\pm0.10$ & $-1.00$ \nl
NGC 5904 (M5) & $-1.40$ &
$1.08\pm0.09$ & $5.78\pm0.04$ & $3.47\pm0.06$ & \phs$0.39$ \nl
NGC 5272 (M3) & $-1.66$ &
$1.19\pm0.10$ & $5.84\pm0.04$ & $3.52\pm0.09$ & \phs$0.07$ \nl 
\tablevspace{0.1in}
NGC 4590 (M68) & $-2.09$ & 
$0.91\pm0.17$ & $6.34\pm0.05$ & $3.41\pm0.05$ & \phs$0.44$ \nl
NGC 5024 (M53) & $-2.04$ &
$1.18\pm0.18$ & \nodata & $3.46\pm0.08$ & \phs$0.76$ \nl
NGC 5053 & $-2.41$ &
$0.95\pm0.25$ & \nodata & $3.44\pm0.08$ & \phs$0.61$ \nl
NGC 5466 & $-2.22$ &
$1.21\pm0.27$ & \nodata & \nodata & \phs$0.51$ \nl
NGC 6341 (M92) & $-2.24$ &
$1.26\pm0.18$ & \nodata & $3.44\pm0.06$ & \phs$0.88$ \nl
NGC 6397 & $-1.91$ & $1.15\pm0.17$ & \nodata &
$3.6\phn\pm0.14$ & \phs$0.93$ (0.69) \nl
NGC 7078 (M15) & $-2.17$ &
$1.23\pm0.21$ & $6.32\pm0.11$ & $3.46\pm0.10$ & \phs$0.72$ \nl
NGC 7099 (M30) & $-2.13$ &
$1.49\pm0.18$ & $6.42\pm0.13$ & $3.59\pm0.06$ & \phs$0.84$ \nl
\enddata
\label{clusttab}
\end{deluxetable}

\begin{deluxetable}{lcccccc}
\tablecolumns{7}
\tablenum{7}
\tablecaption{Bright Star Populations and Population Ratios}
\tablehead{\colhead{Sample} & \colhead{BHB} & \colhead{RR Lyr} &
\colhead{RHB} & \colhead{Total HB} & \colhead{AGB} & \colhead{RGB}\\
 & \colhead{$R$} & \colhead{$R'$} & \colhead{$R_{1}$} & \colhead{$R_{2}$} &
\colhead{$R_{HB}$} & }
\scriptsize
\startdata
$r < 30''$ & 57 & 2: & 1 & 60 & 3 & 46 \nl
& $1.30\pm0.26$ & $1.22\pm0.24$ & $0.07\pm0.04$ & $0.05\pm0.03$ &
$0.93\pm0.04$ & \nl
$30'' < r < 100''$ & 59 & 4 & 2 & 65 & 4 & 48 \nl
& $1.35\pm0.27$ & $1.25\pm0.24$ & $0.08\pm0.04$ & $0.06\pm0.03$ &
$0.88\pm0.05$ & \nl
$r > 100''$ & 65 & 3 & 9 & 77 & 4 & 42 \nl
& $1.83\pm0.36$ & $1.67\pm0.32$ & $0.10\pm0.05$ & $0.05\pm0.03$ &
$0.73\pm0.07$ & \nl
Total & 181 & 9 & 12 & 202 & 11 & 136 \nl
& $1.49\pm0.18$ & $1.37\pm0.16$ & $0.08\pm0.03$ & $0.05\pm0.02$ &
$0.84\pm0.04$ & \nl
YGSB & 51 & 1: & 1 & 53 & 1 & 34 \nl
& $1.56\pm0.35$ & \nodata & \nodata & \nodata & \nodata & \nl
\enddata
\label{poptab}
\end{deluxetable}

\begin{deluxetable}{ccccc}
\tablecolumns{5}
\tablewidth{0pc}
\tablenum{A1}
\tablecaption{Photometric Transformation Equation Coefficients}
\tablehead{\colhead{Band} & \colhead{Zero Point} & \colhead{Extinction
Coeffs.} & \colhead{Color Term} & \colhead{Order}}
\startdata
\multicolumn{5}{c}{Primary Standard Calibration: CTIO 1.5~m} \nl
$V$ & $2.577\pm0.001$ & $0.1004\pm0.0046$ & \phs$0.0103\pm0.0026$ & 1 \nl
& & & $-0.0839\pm0.0061$ & 2 \nl
& & & \phs$0.0408\pm0.0061$ & 3 \nl
$I$ & $3.311\pm0.002$ & $0.0532\pm0.0062$ & $0.0002\pm0.0020$ & 1 \nl
\multicolumn{5}{c}{Secondary Standard Calibration: CTIO 1.5~m} \nl
$V$ & \nodata & \nodata & $-0.0361\pm0.0010$ & 1 \nl
$I$ & \nodata & \nodata & $-0.0045\pm0.0018$ & 1 \nl
\multicolumn{5}{c}{Calibration: CTIO 4~m} \nl
$V$ & \nodata & \nodata & $-0.0727\pm0.0068$ & 1 \nl
 &  &  & $0.0514\pm0.0070$ & 2 \nl
$I$ & \nodata & \nodata & $-0.1481\pm0.0095$ & 1 \nl
 &  &  & $0.1570\pm0.0098$ & 2 \nl
\enddata
\label{coeftab}
\end{deluxetable}

\begin{deluxetable}{ccccccccc}
\tablecolumns{9}
\tablewidth{0pc}
\tablenum{A2}
\tablecaption{Average Residuals for Photometry Comparisons}
\tablehead{\colhead{Sample 1} & \colhead{Sample 2} & \colhead{$V$} &
\colhead{$\sigma_{V}$} & \colhead{$I$} & \colhead{$\sigma_{I}$} &
\colhead{$(V-I)$} & \colhead{$\sigma_{V-I}$} & \colhead{$N$}}
\startdata
1.5 m & Landolt & 0.0005 & 0.0006 & $-0.0004$ & 0.0119 &
$-0.0005$ & 0.0010 & 27 \nl
PSF & Aperture & $-0.0005$ & 0.0204 & $-0.0014$ & 0.0198 &
$0.0004$ & 0.0252 & 118 \nl
4 m & 1.5 m & $-0.0064$ & 0.0447 & $-0.0125$ & 0.0635 &
$0.0002$ & 0.0515 & 248 \nl
4 m + 1.5 m & Bolte 1987 (s) & $-0.0201$ & 0.1067 & \nodata & \nodata & 59 \nl
4 m + 1.5 m & Bolte 1987 (l) & $-0.0554$ & 0.0928 & \nodata & \nodata & 401 \nl
4 m + 1.5 m & RFV & $-0.0738$ & 0.1419 & \nodata & \nodata &
1374 \nl
4 m + 1.5 m & Samus et al. 1995 & $0.0439$ & 0.1449 &
$0.0463$ & 0.1771 & $0.0030$ & 0.1139 & 255 \nl
4 m + 1.5 m & Bergbusch 1996 & $-0.1101$ & 0.0611 & \nodata & \nodata & 316 \nl
\enddata
\label{restab}
\end{deluxetable}
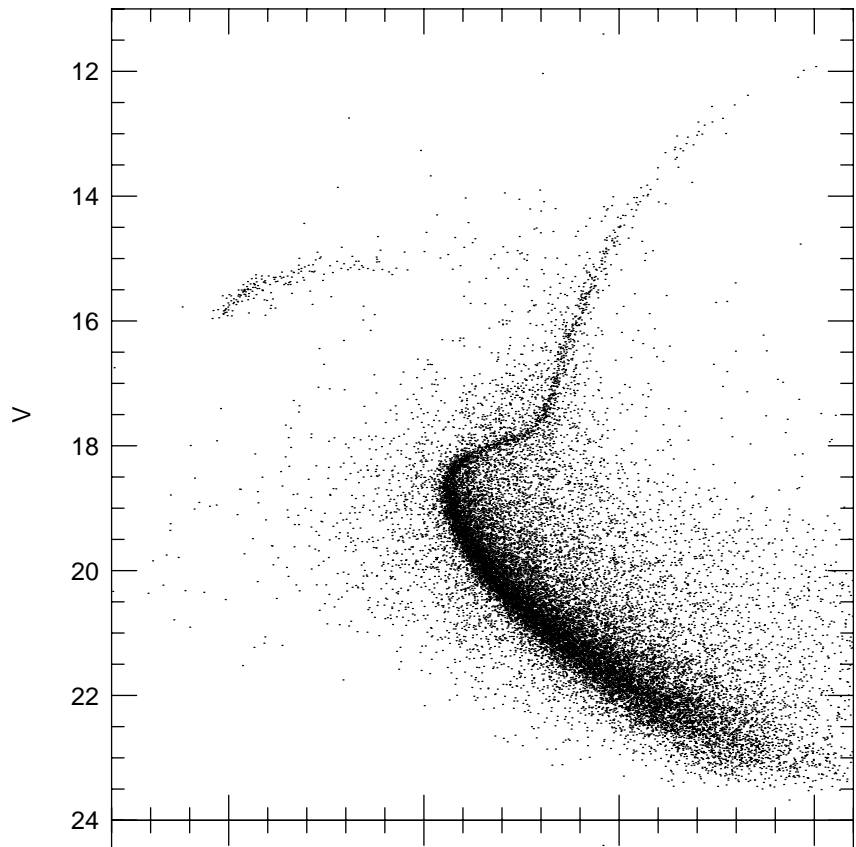
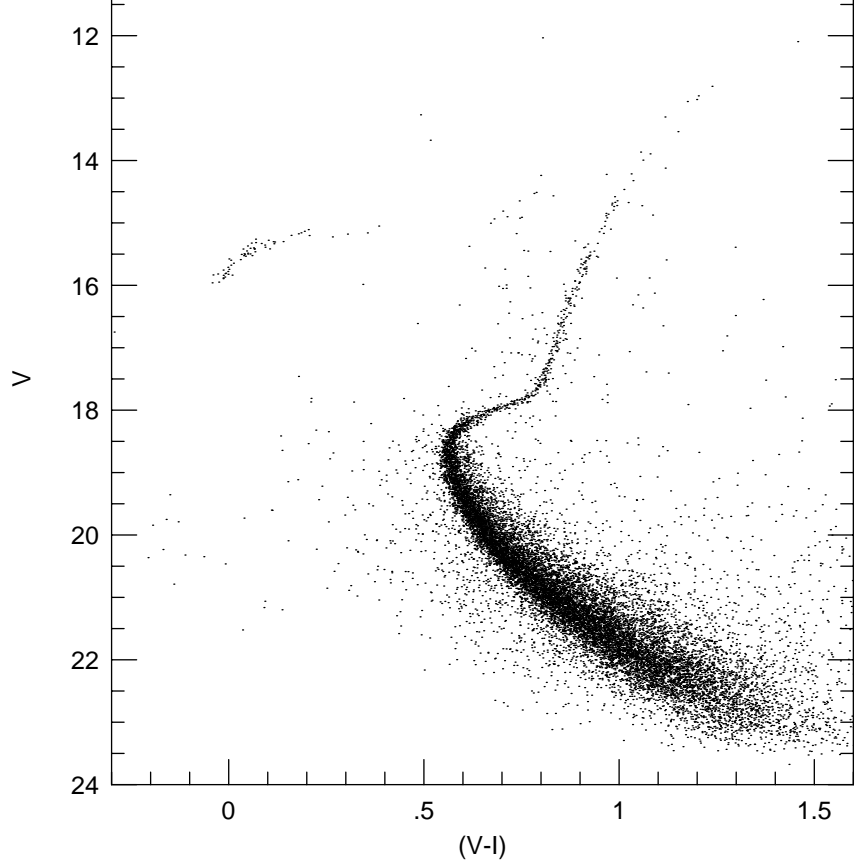
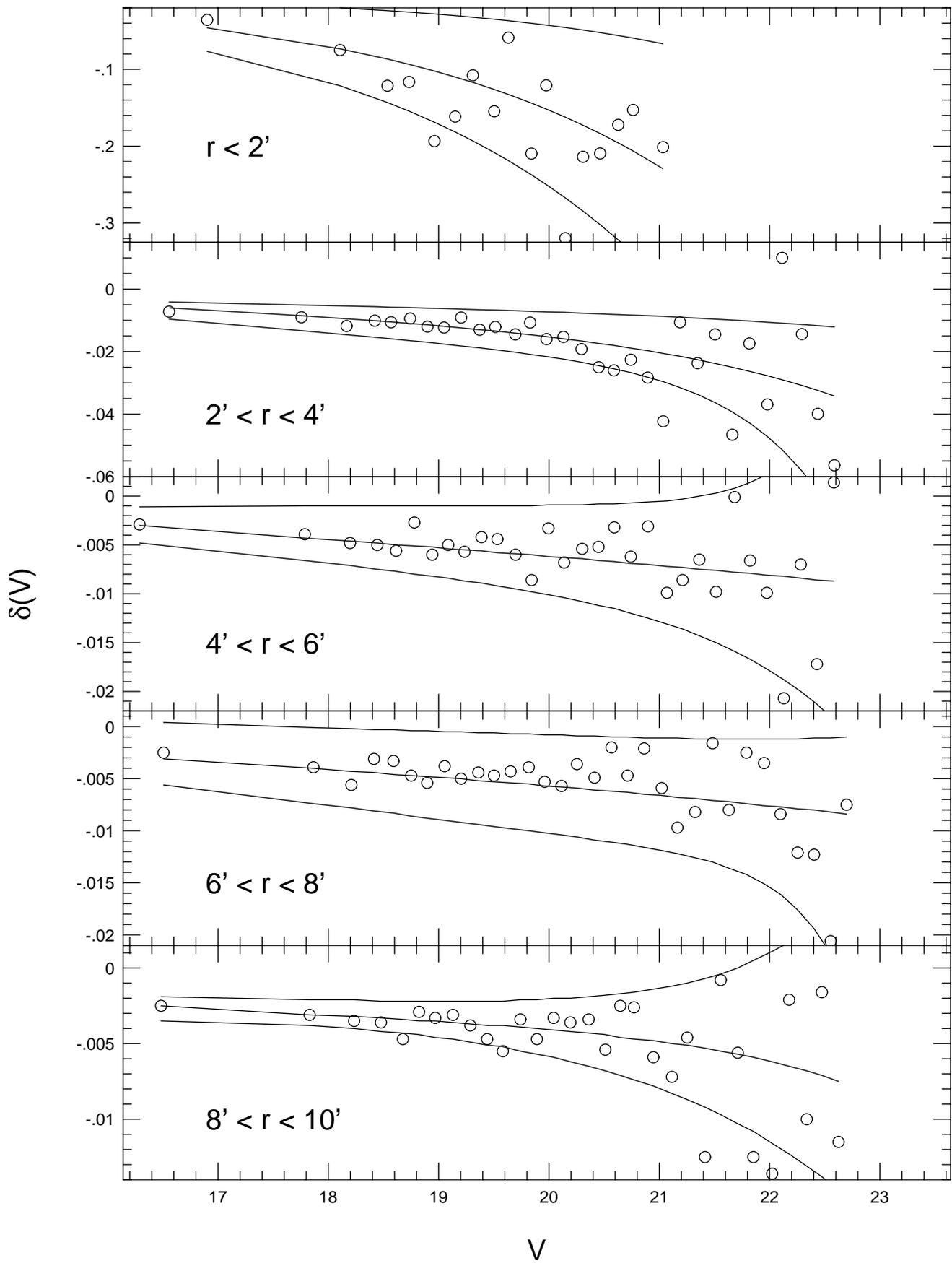
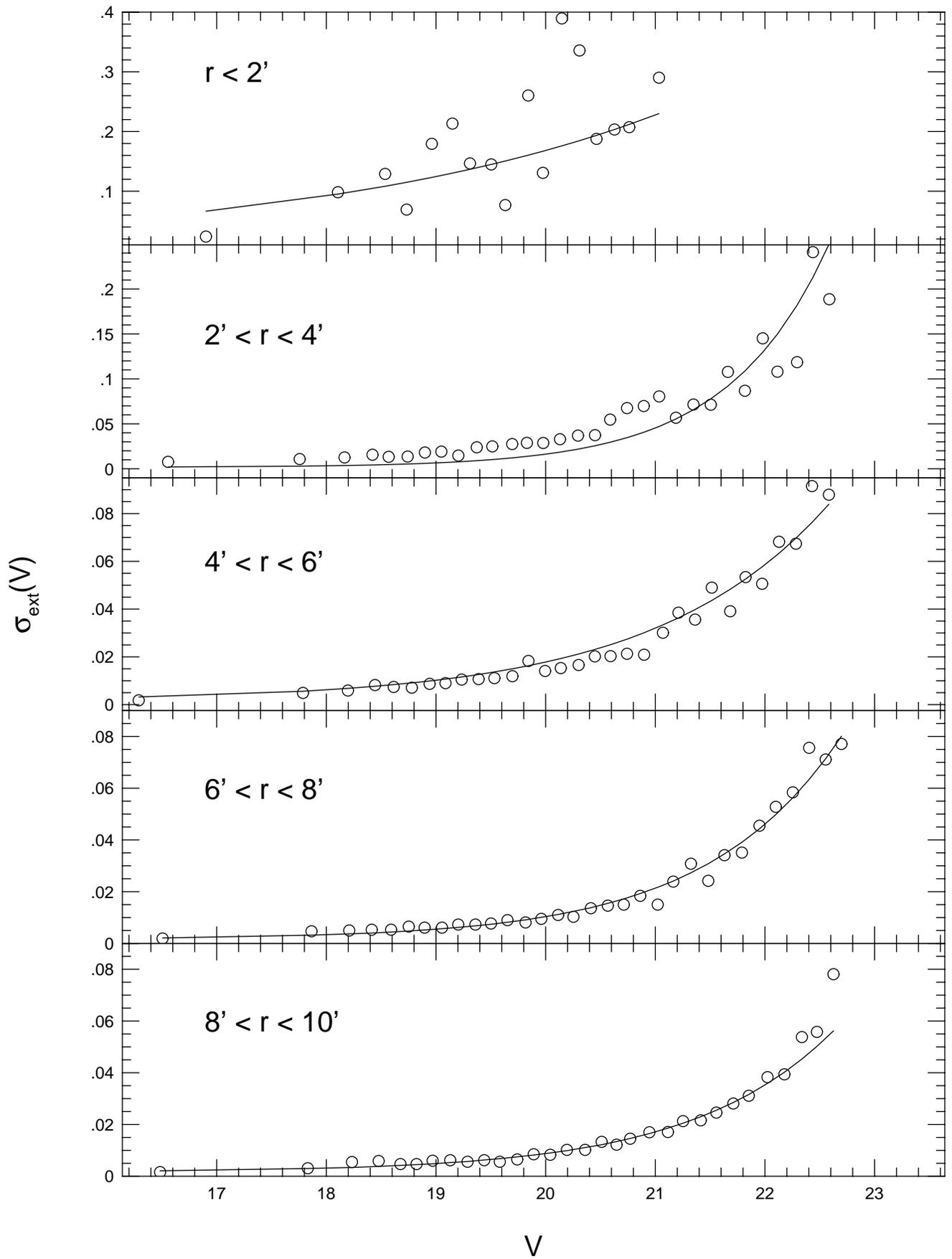
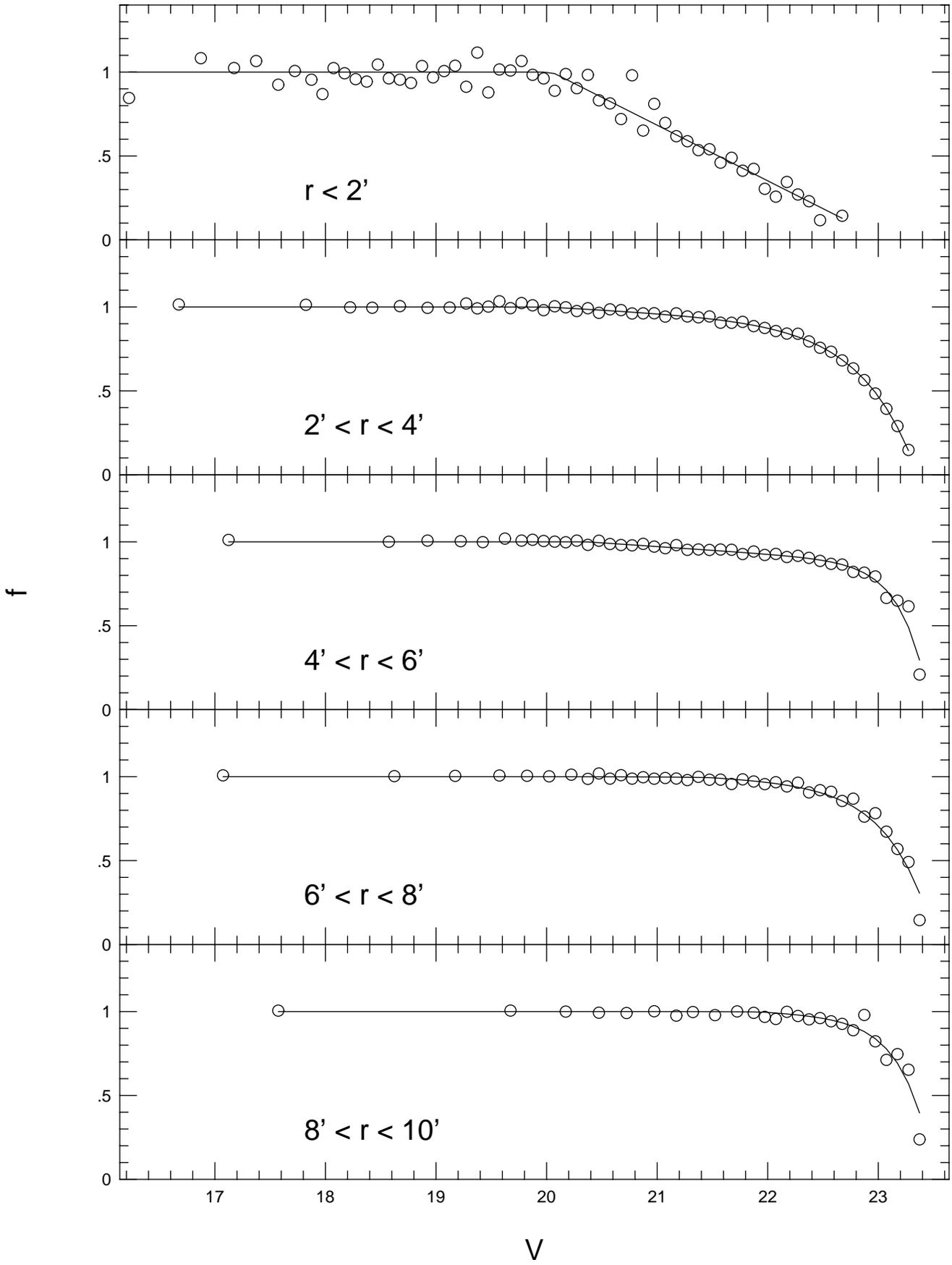
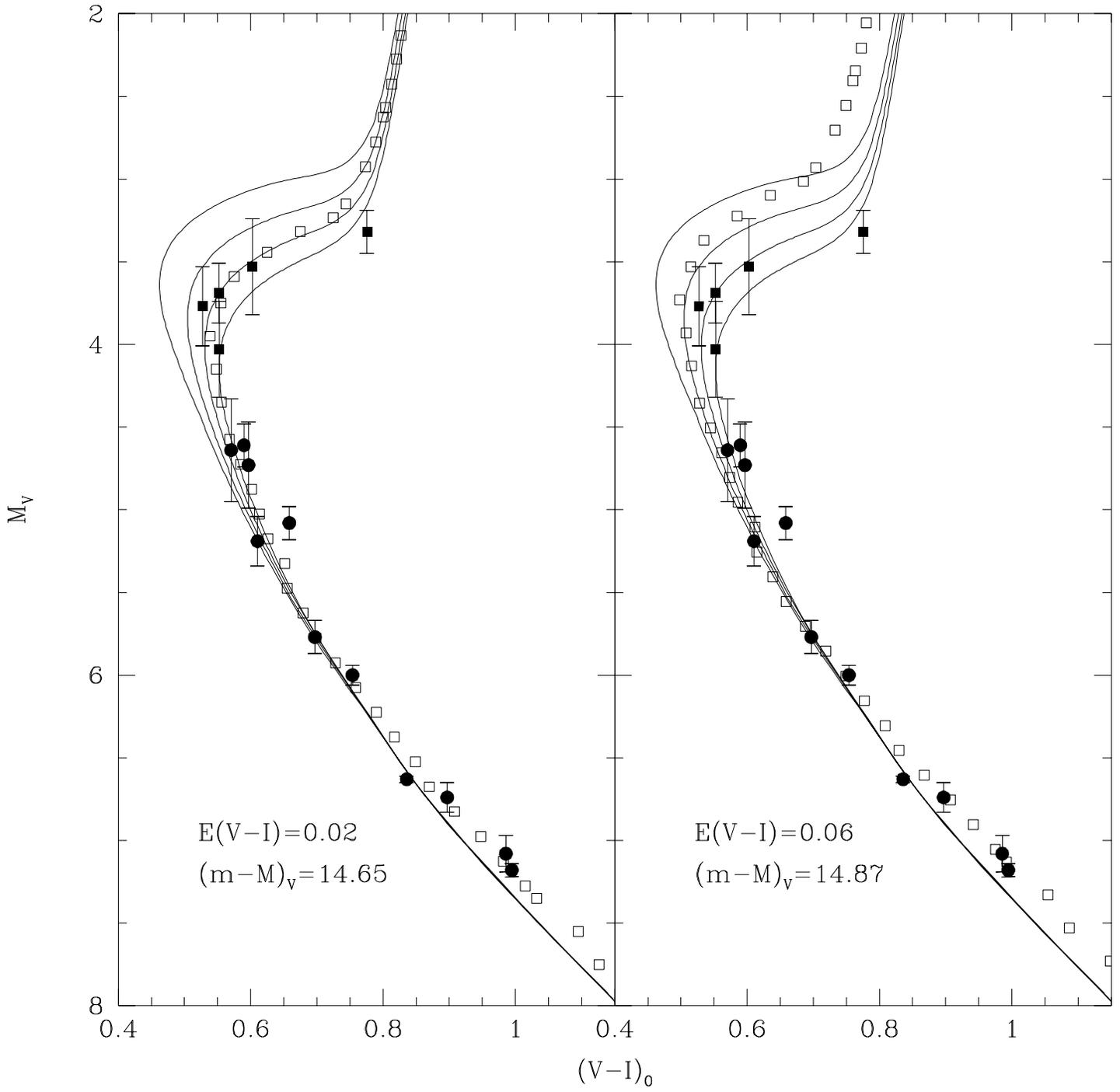
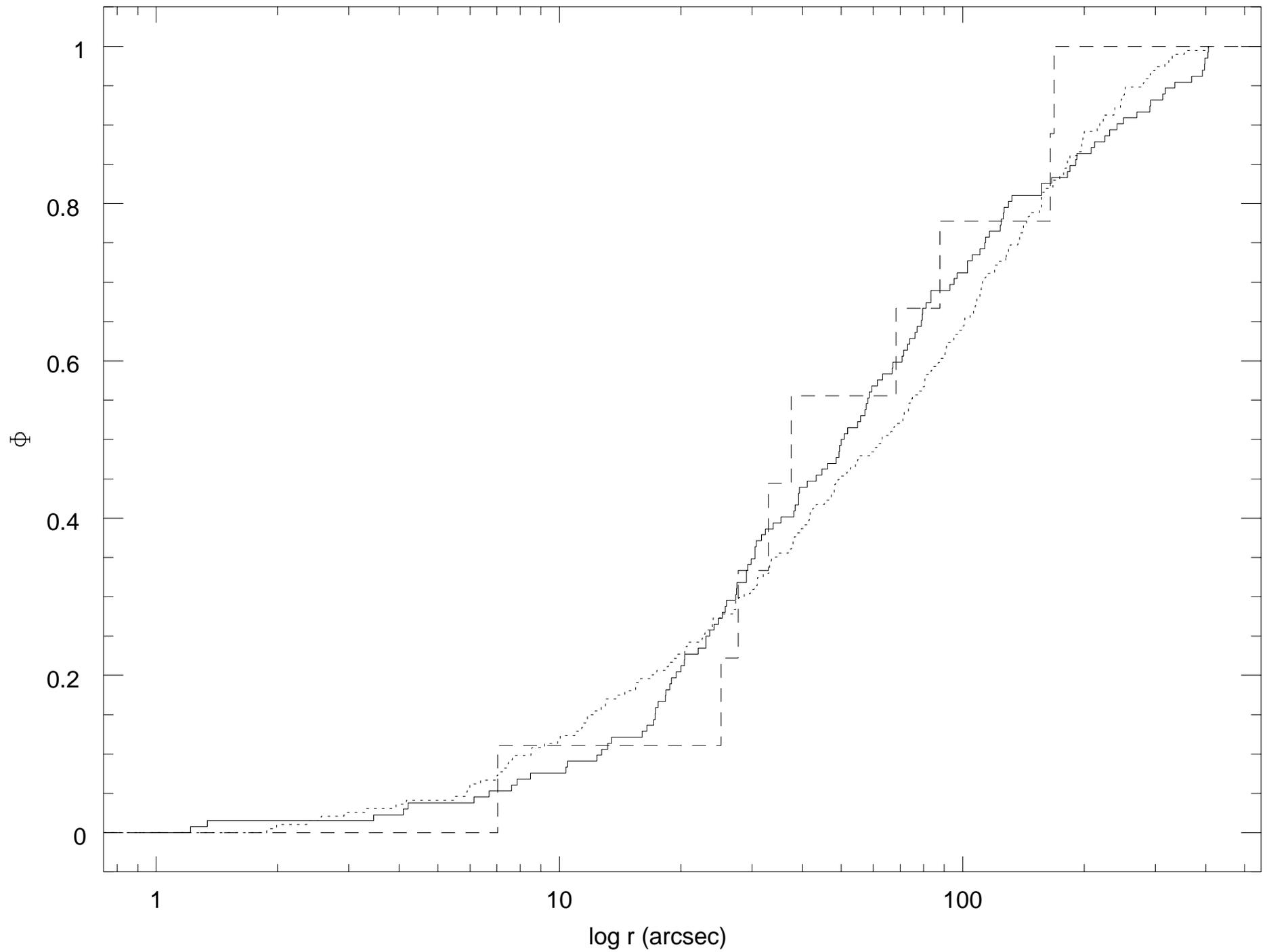
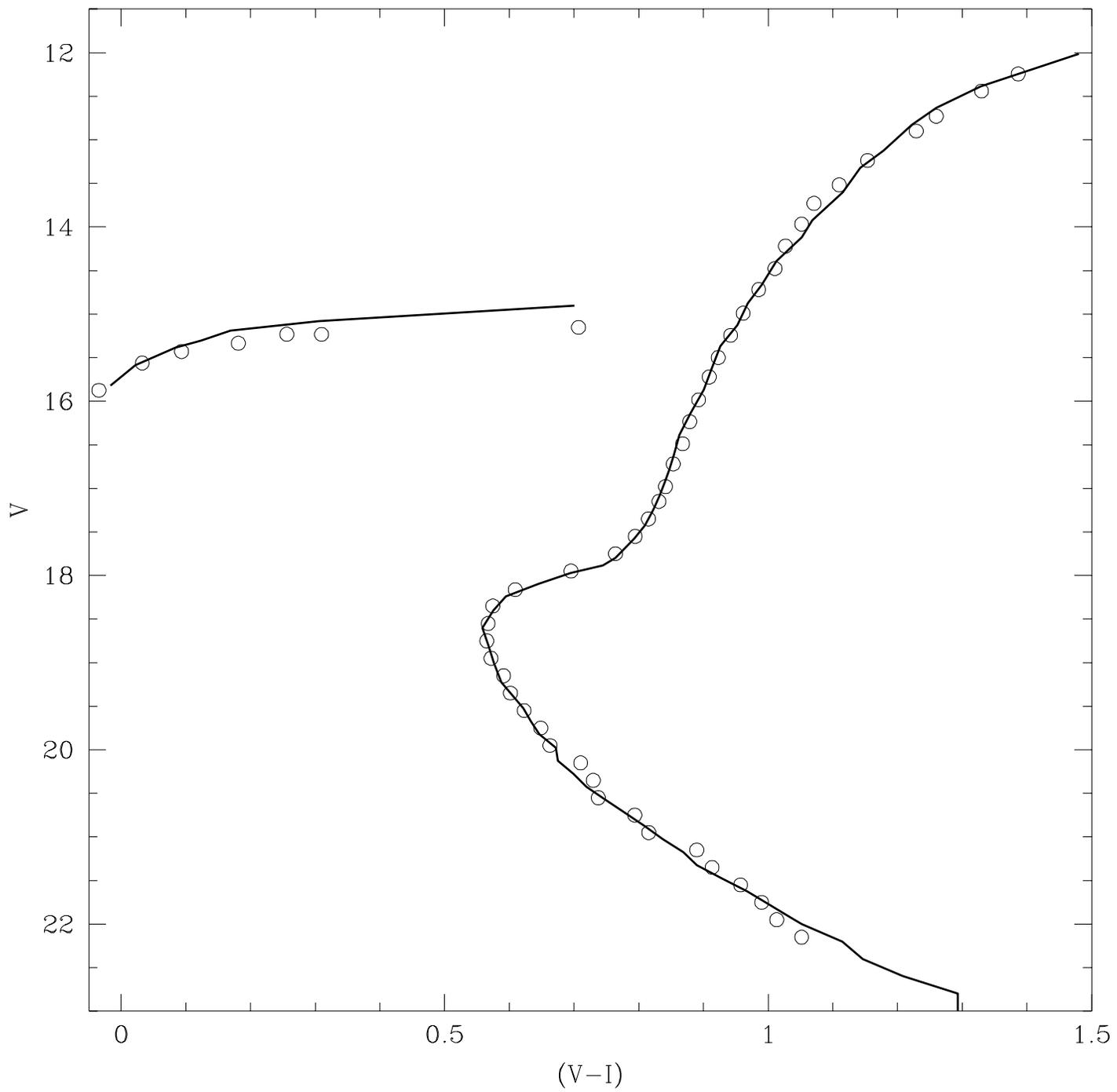
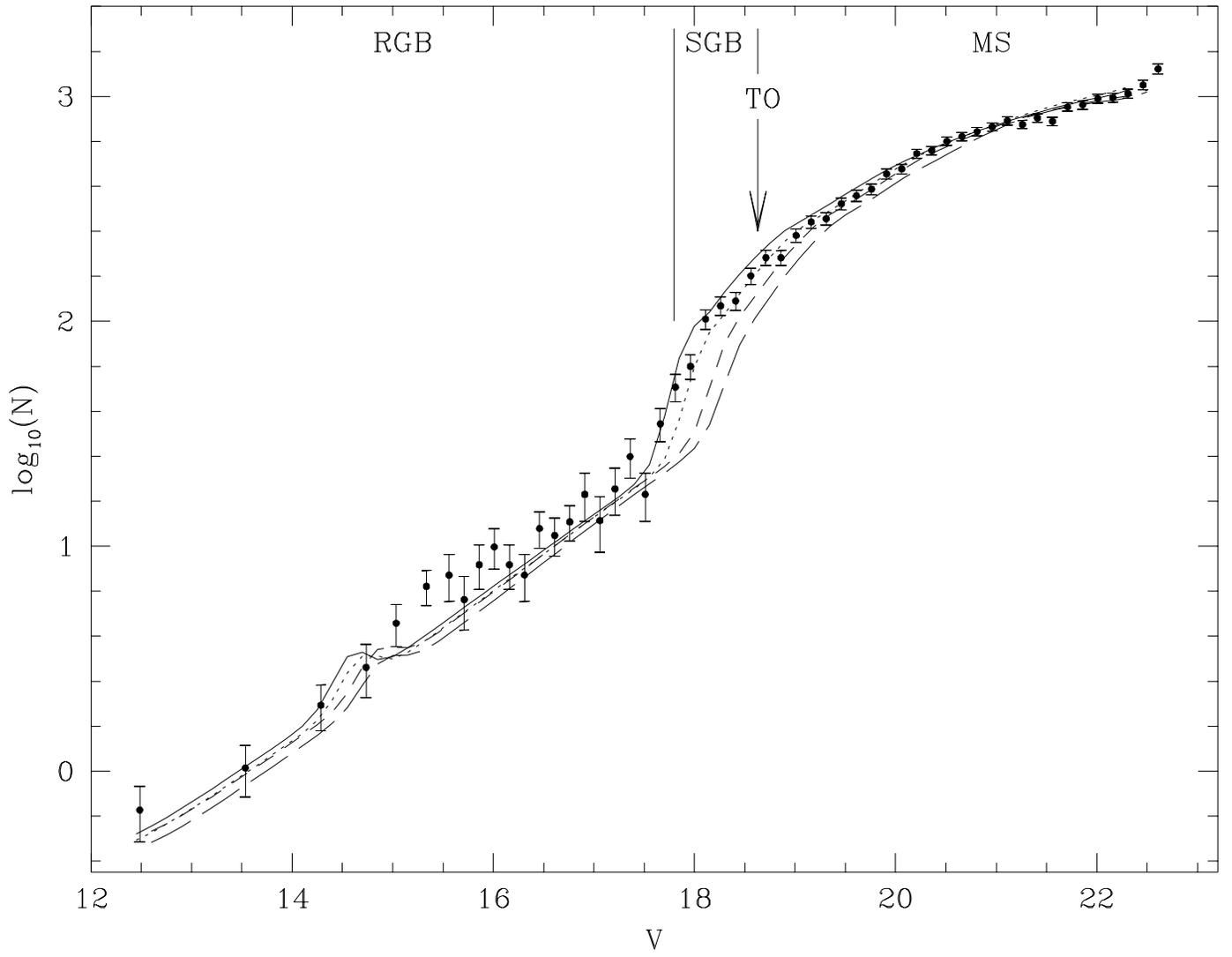
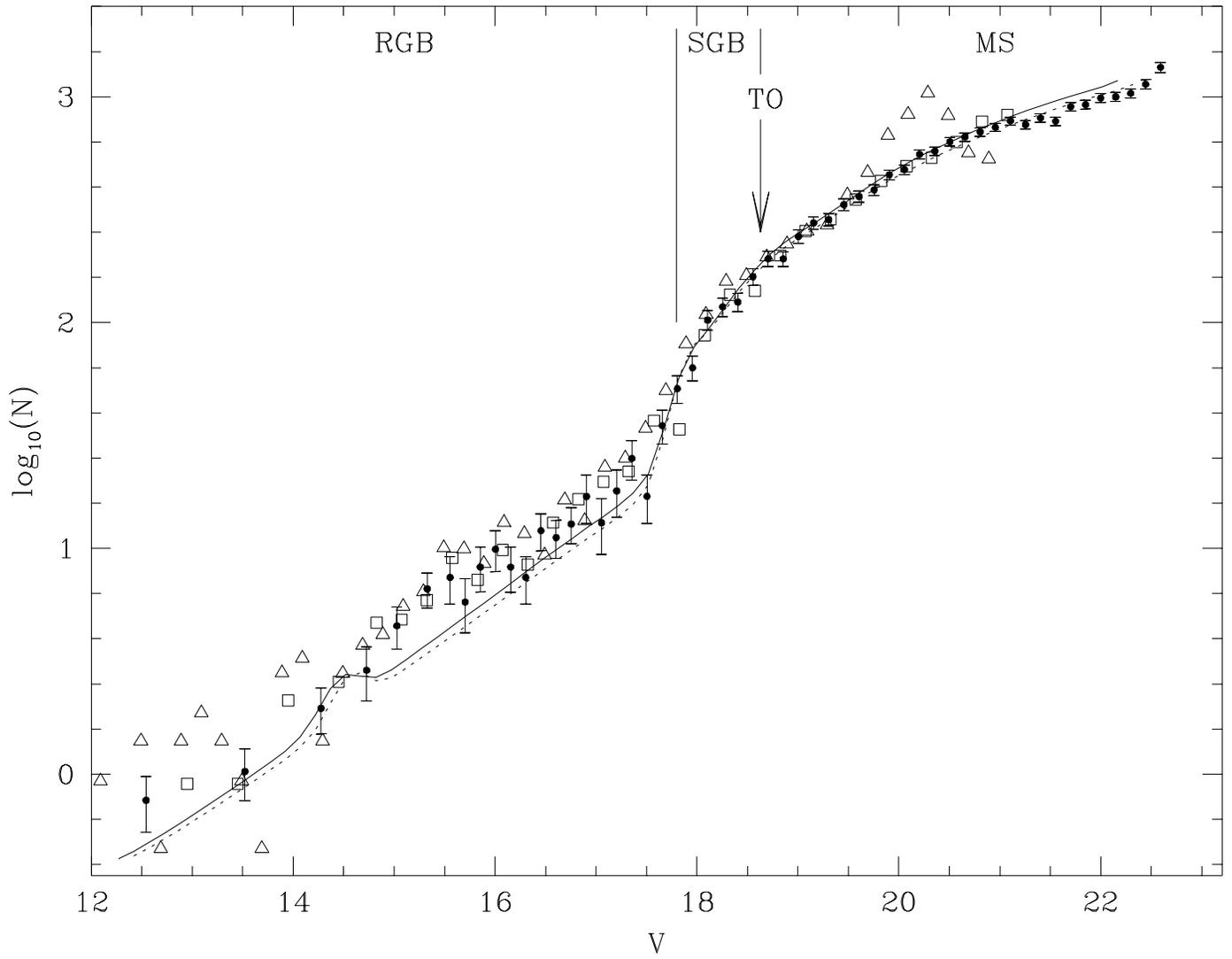
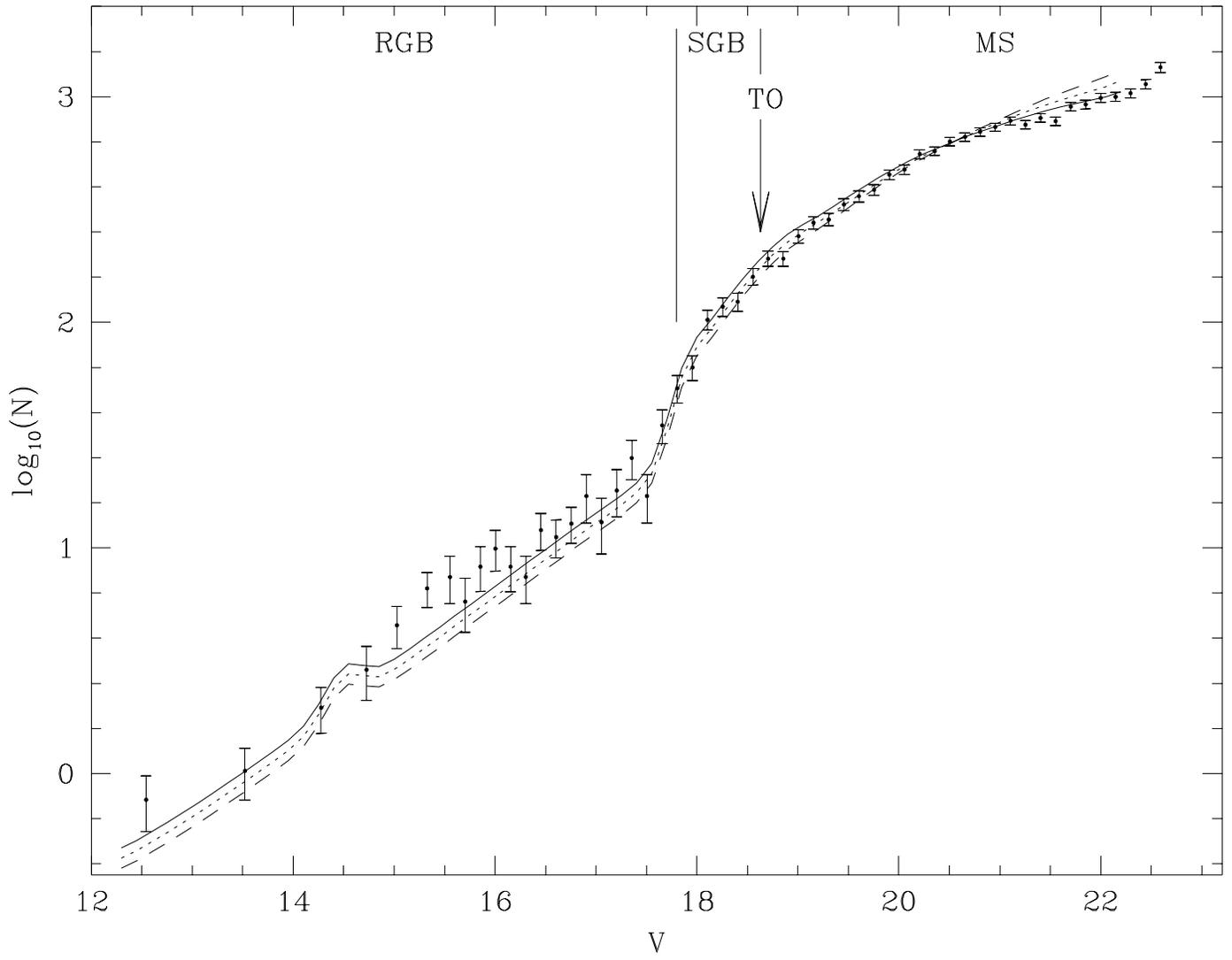
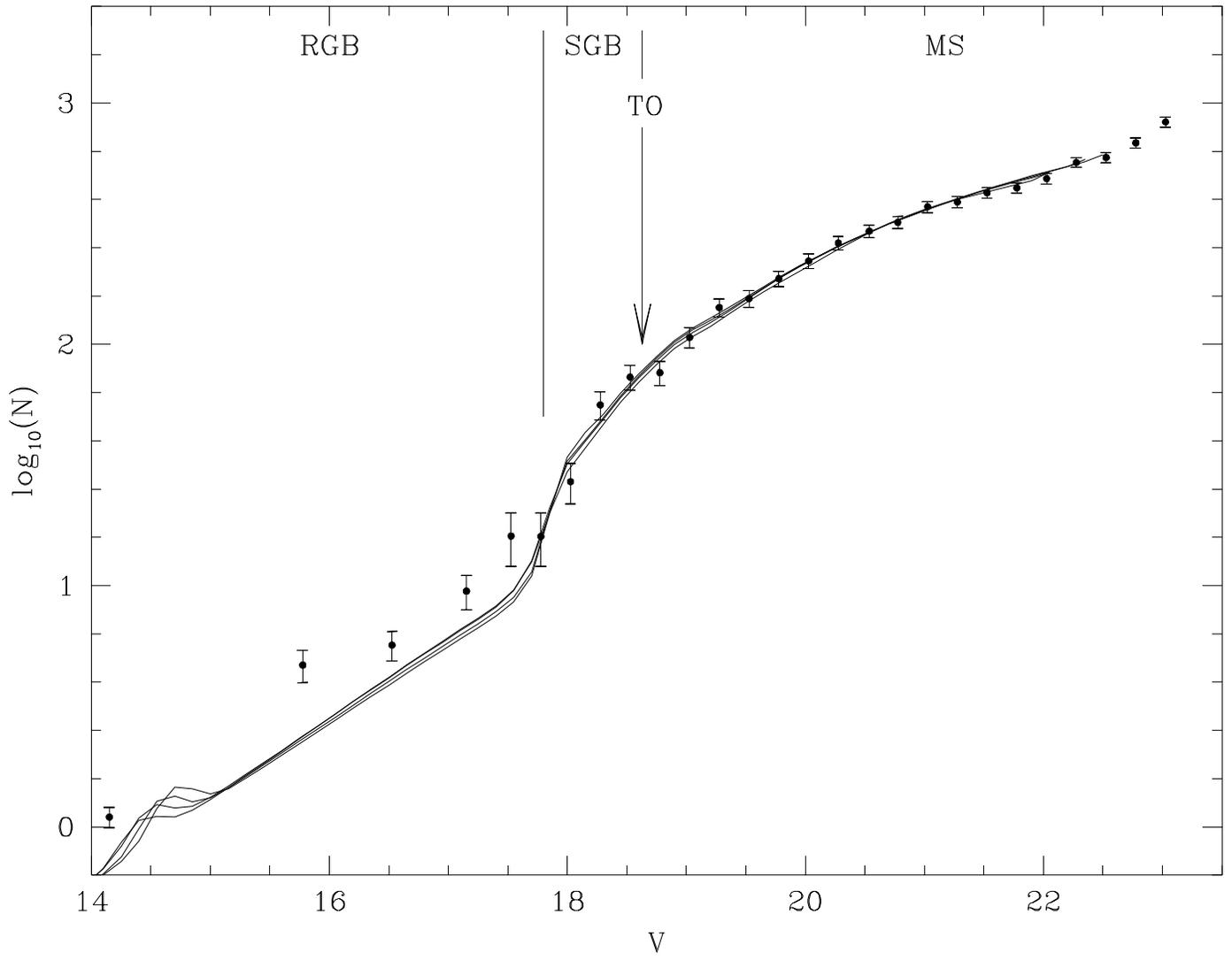
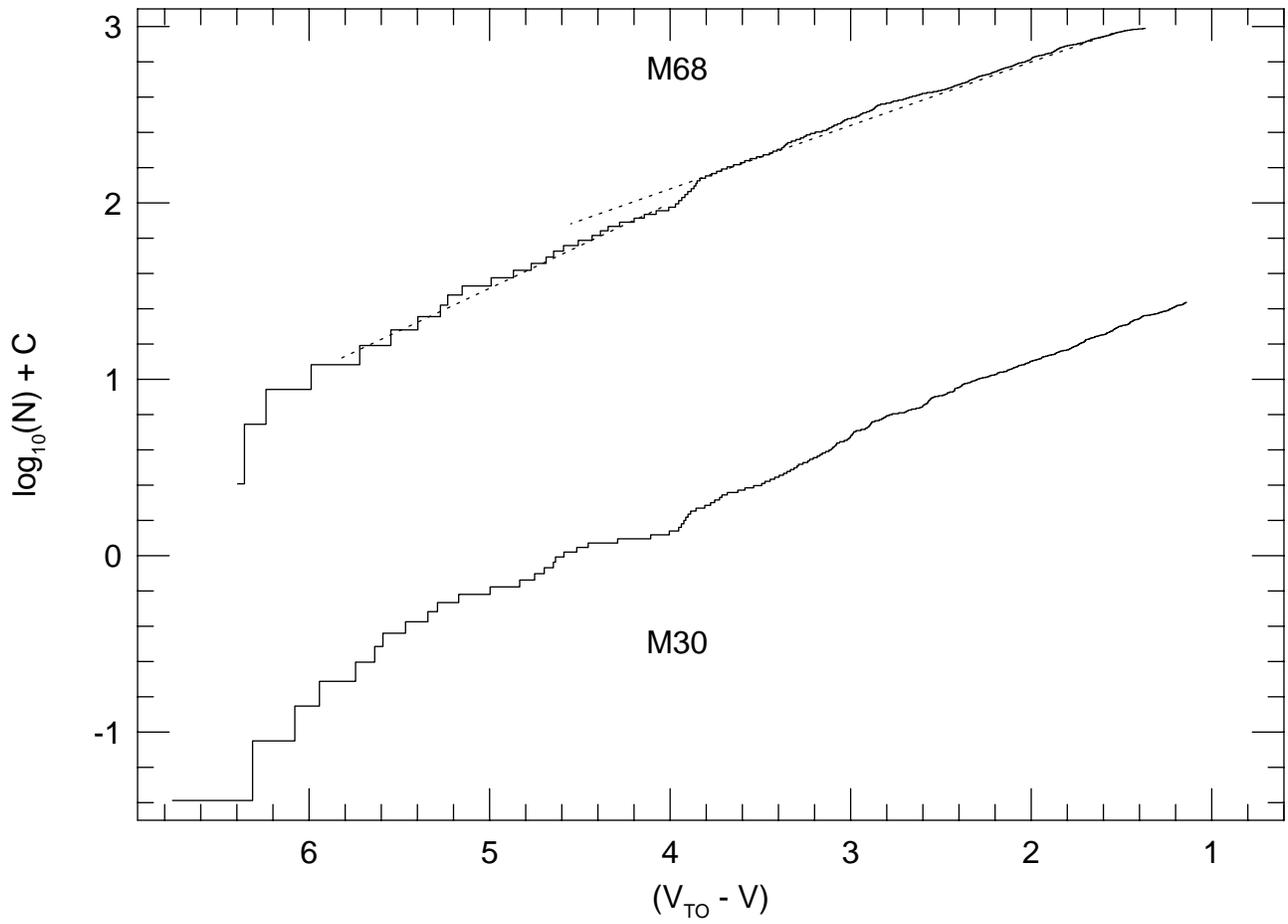
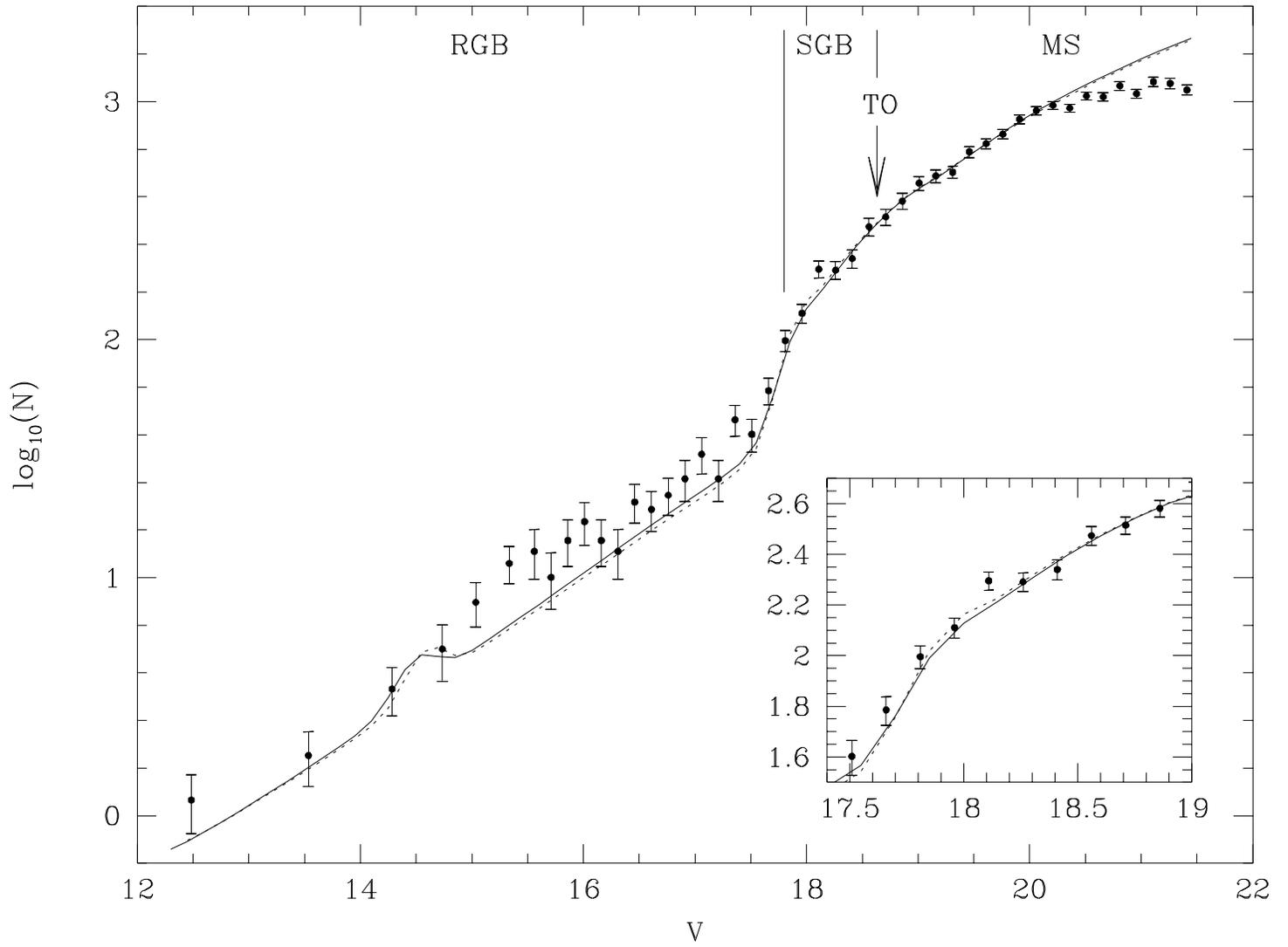
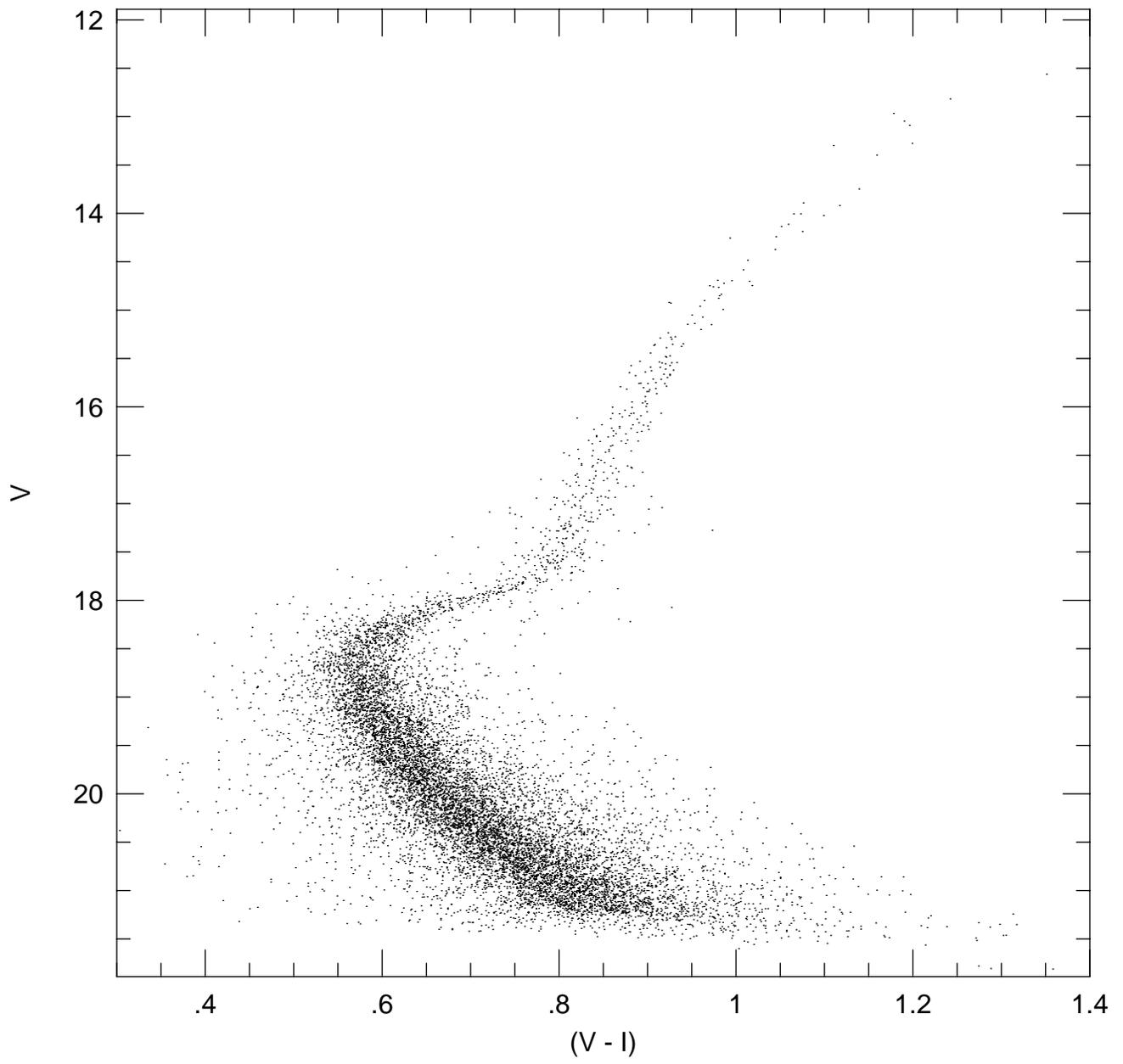
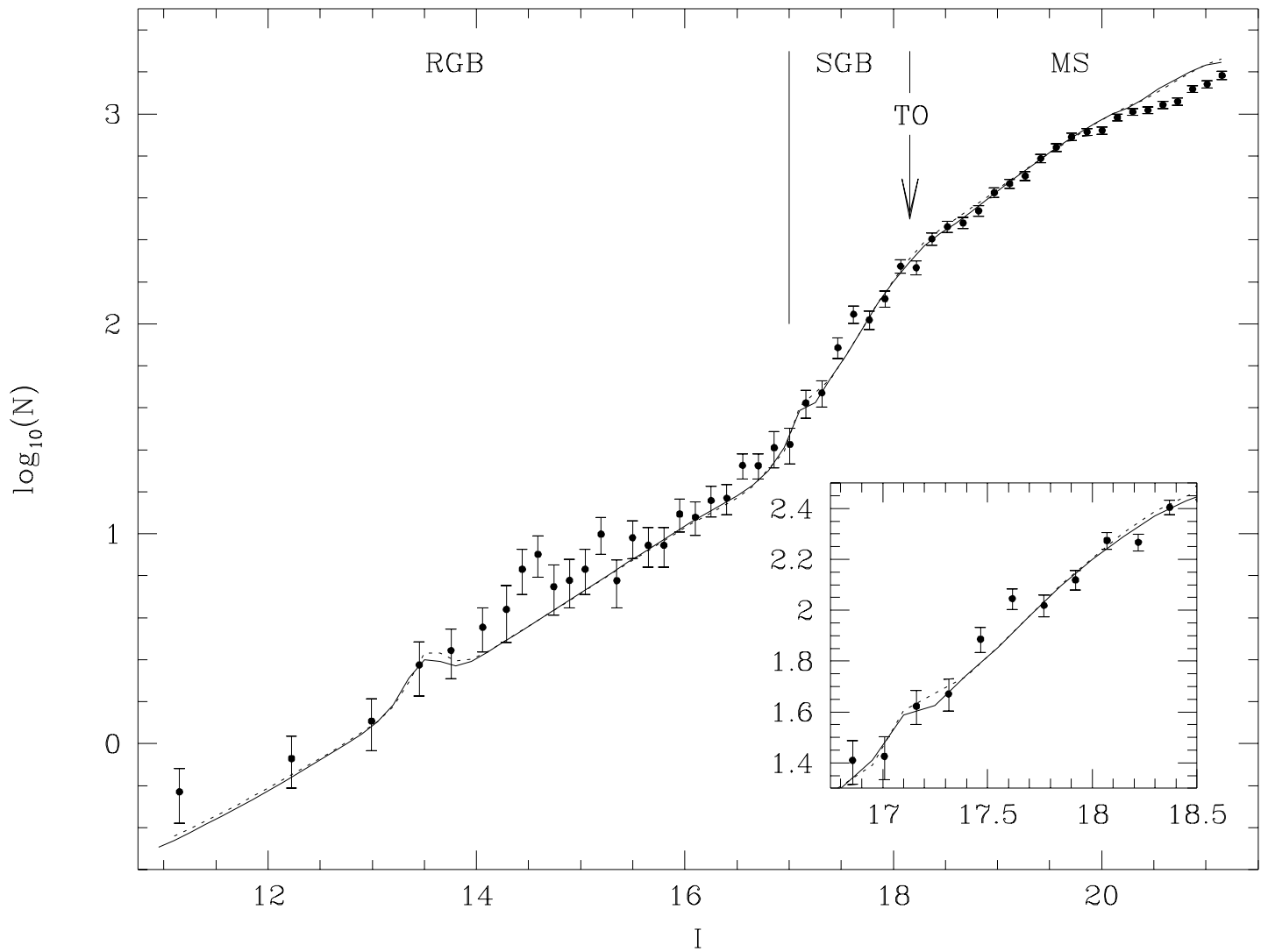
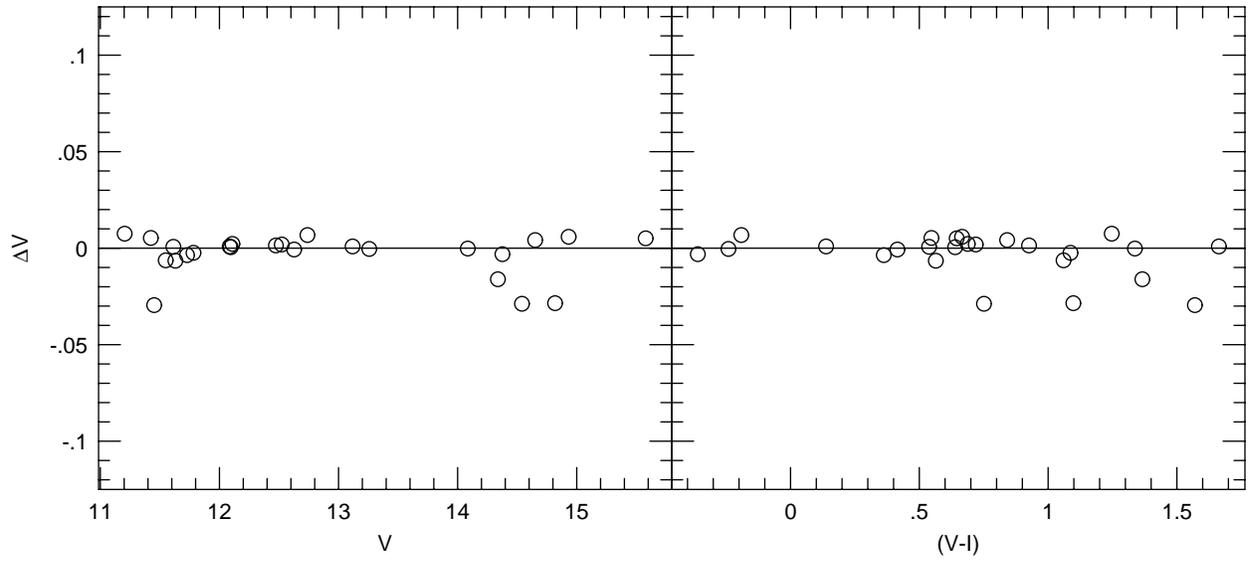
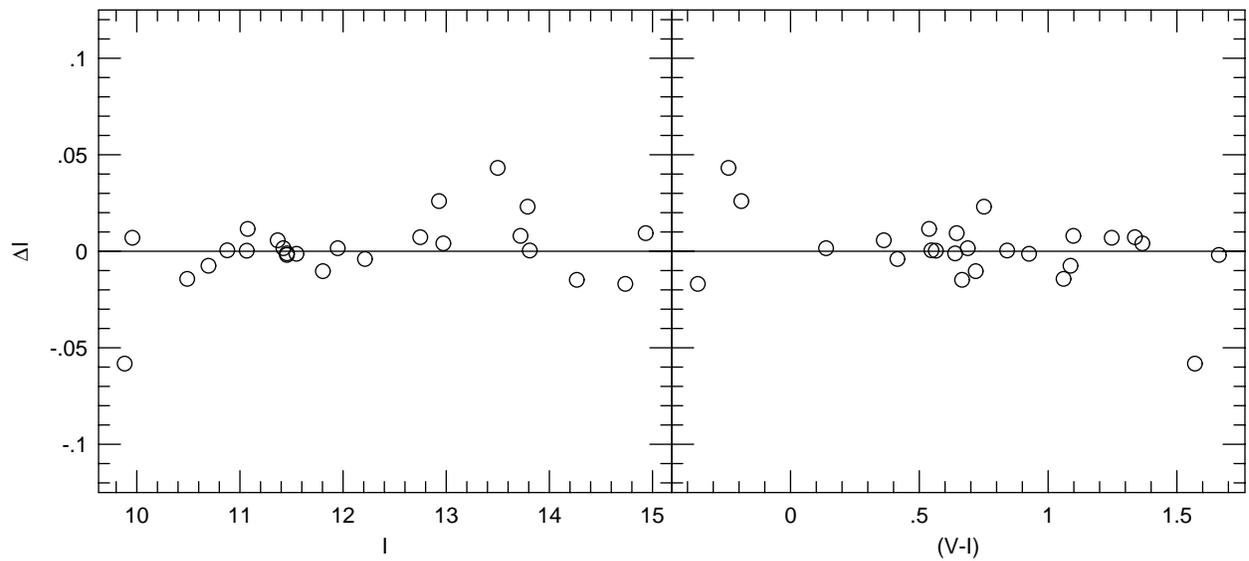
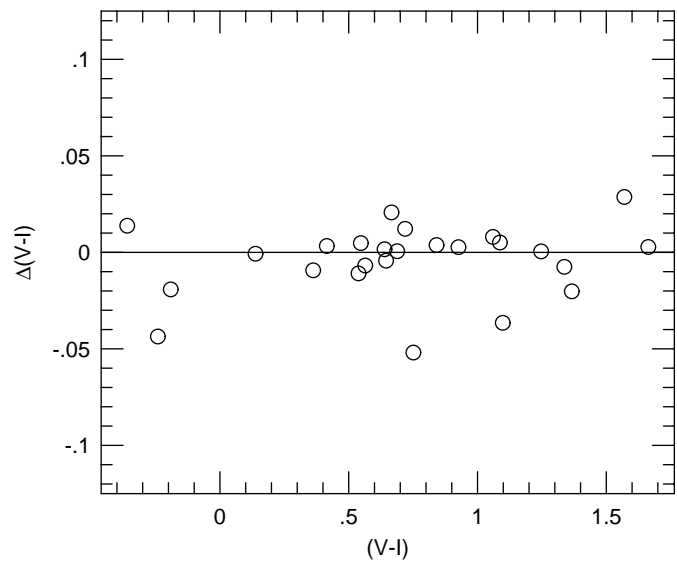
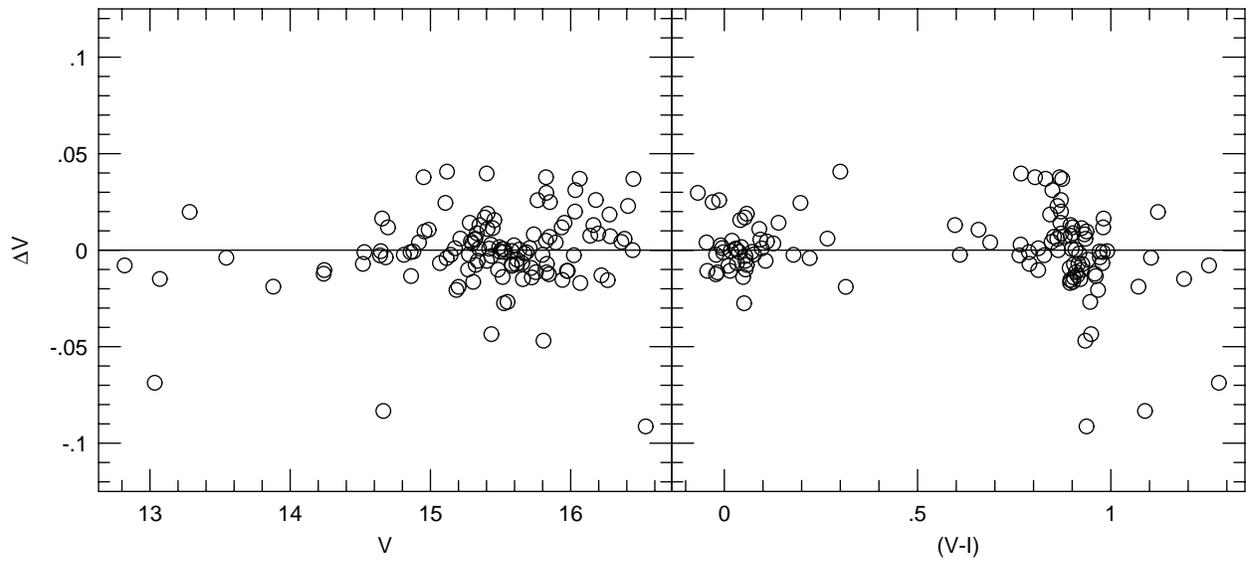
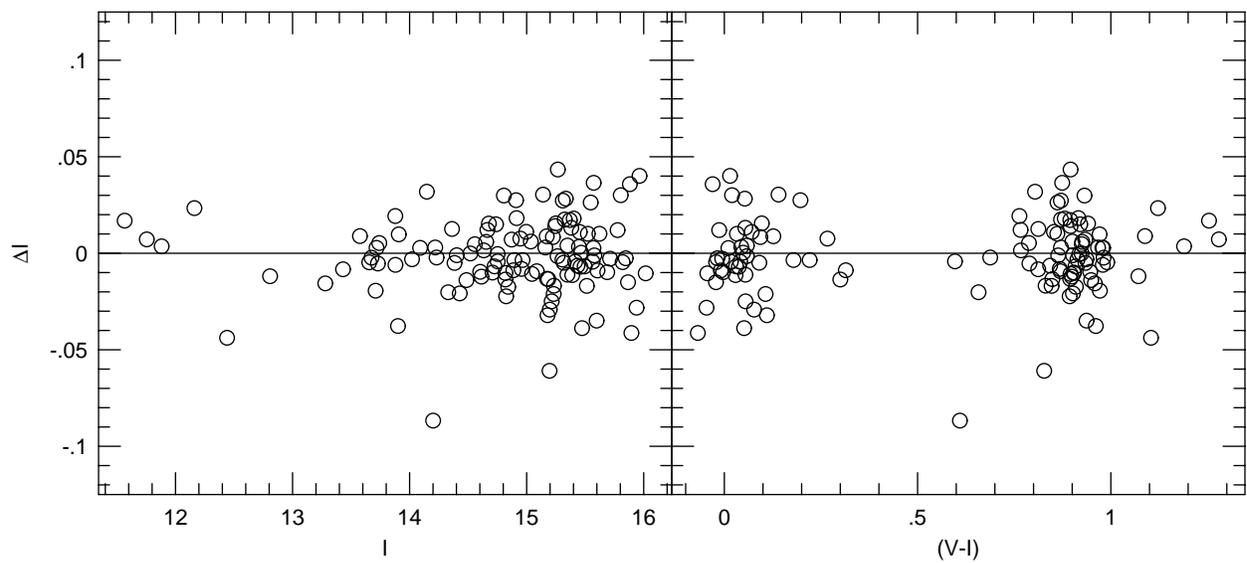
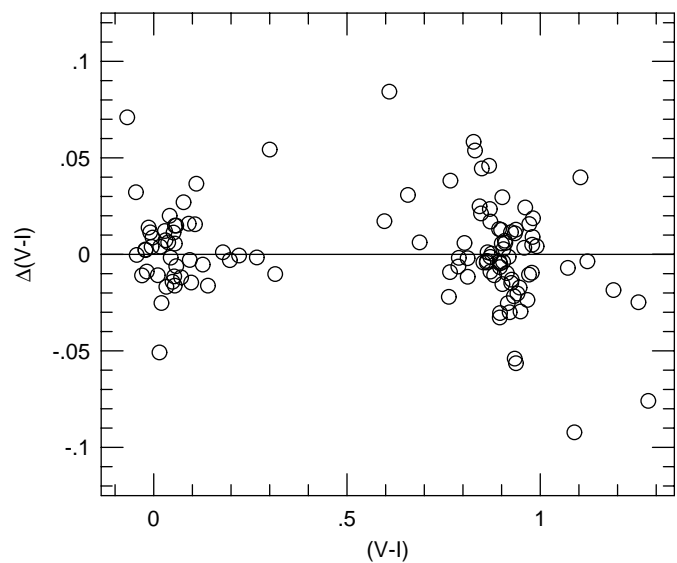
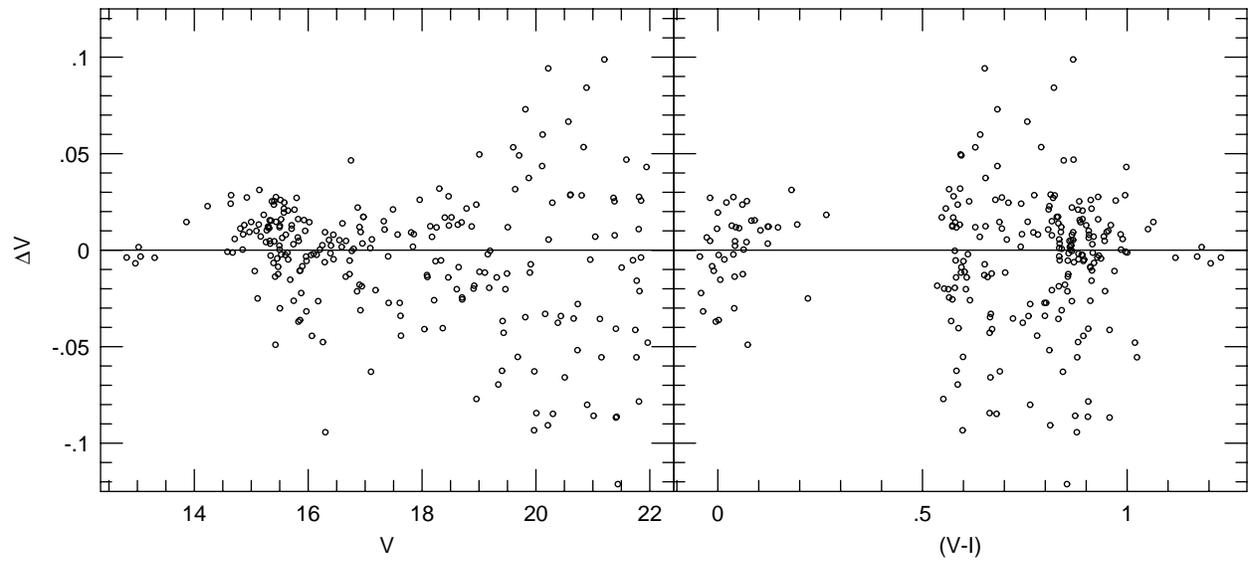
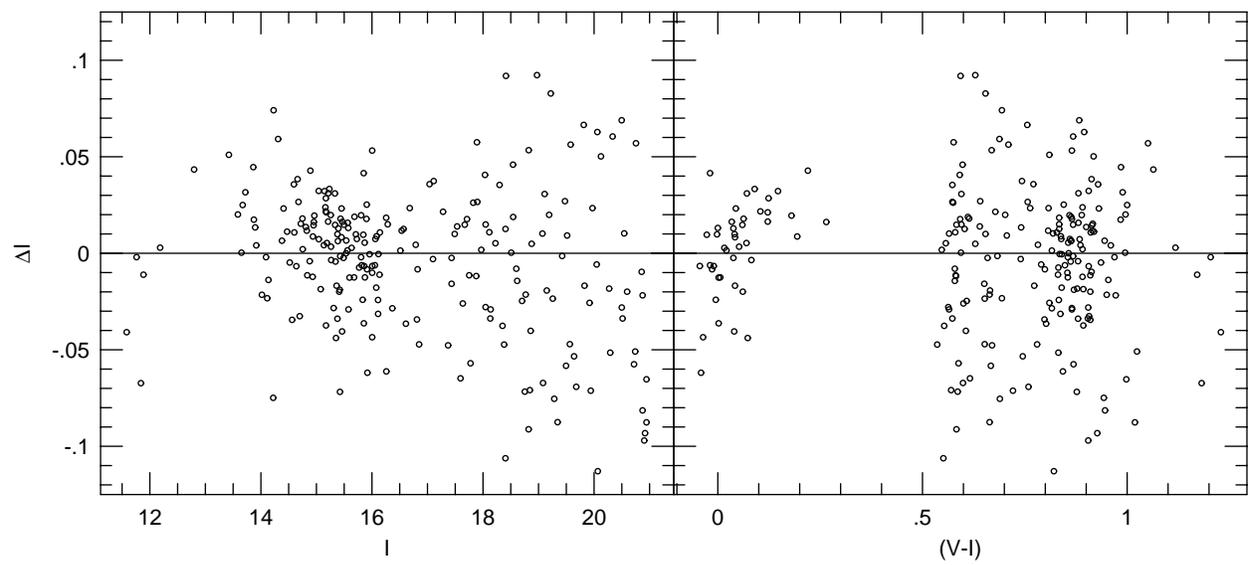
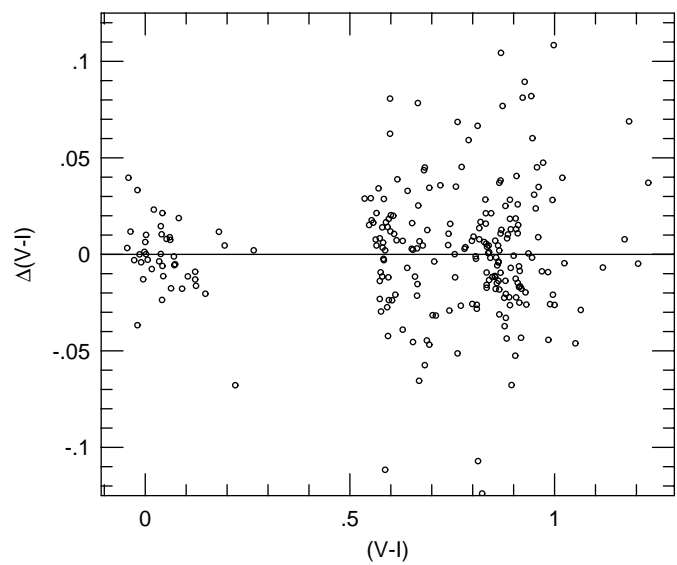
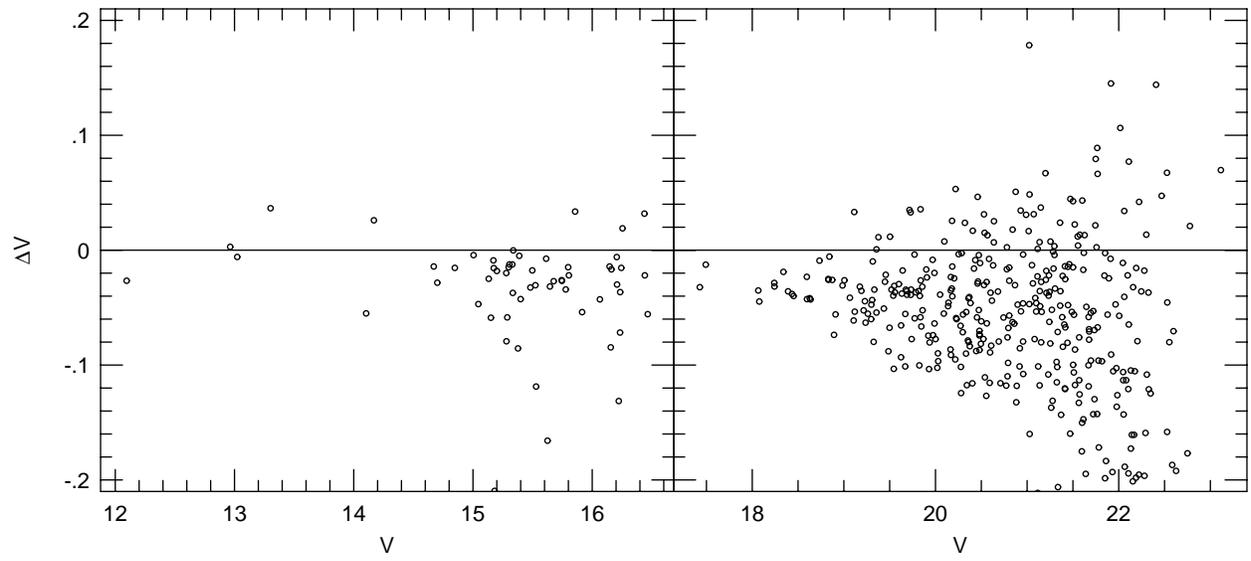
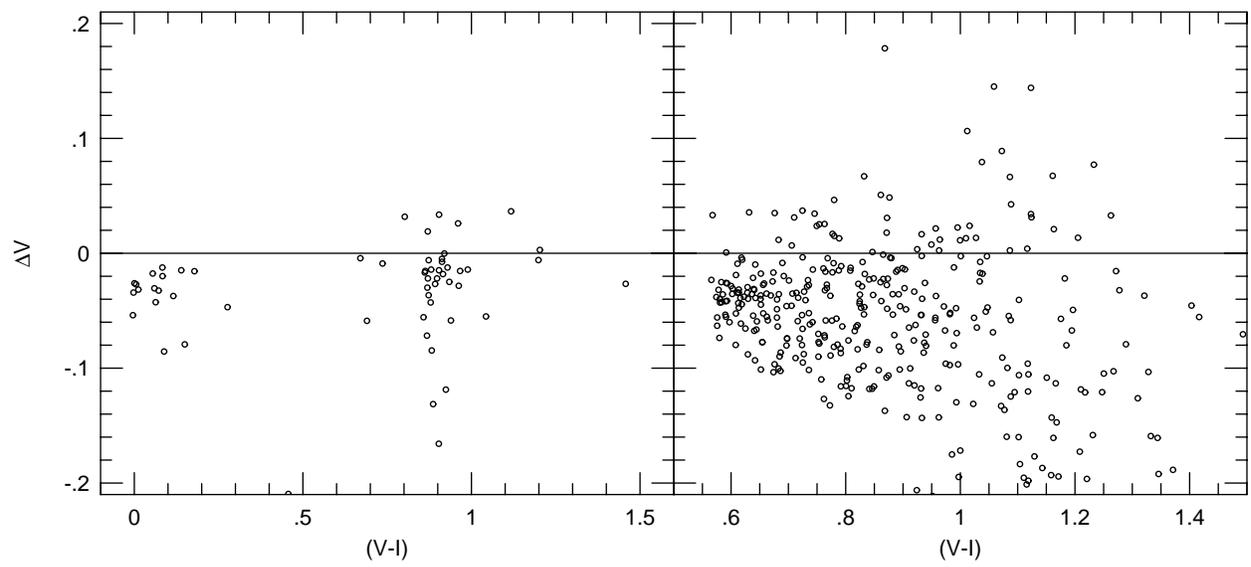
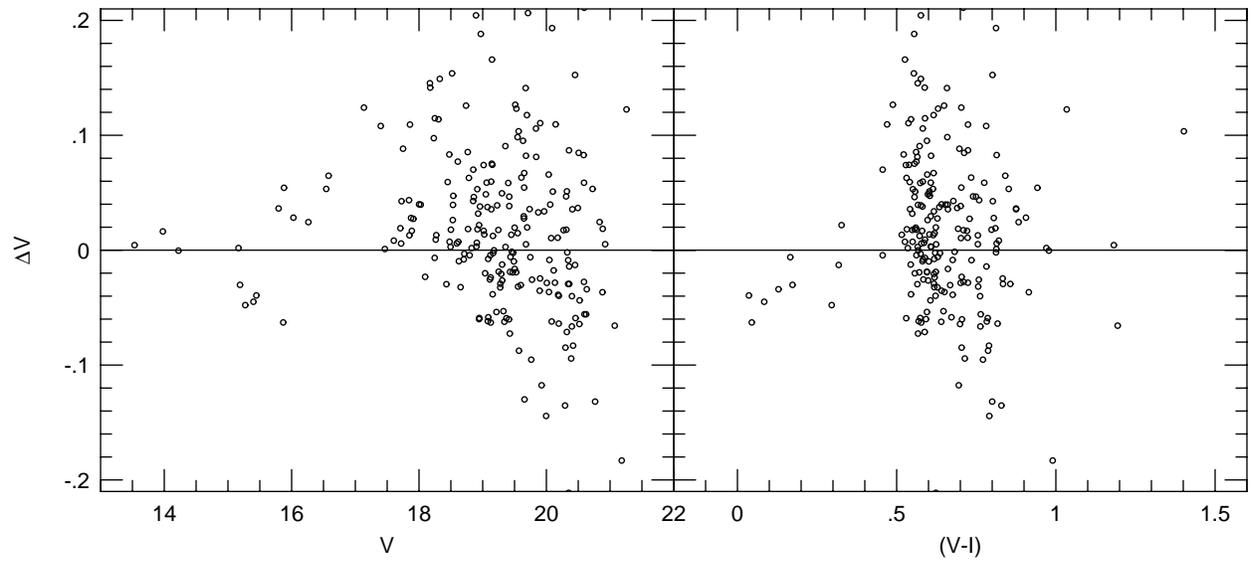
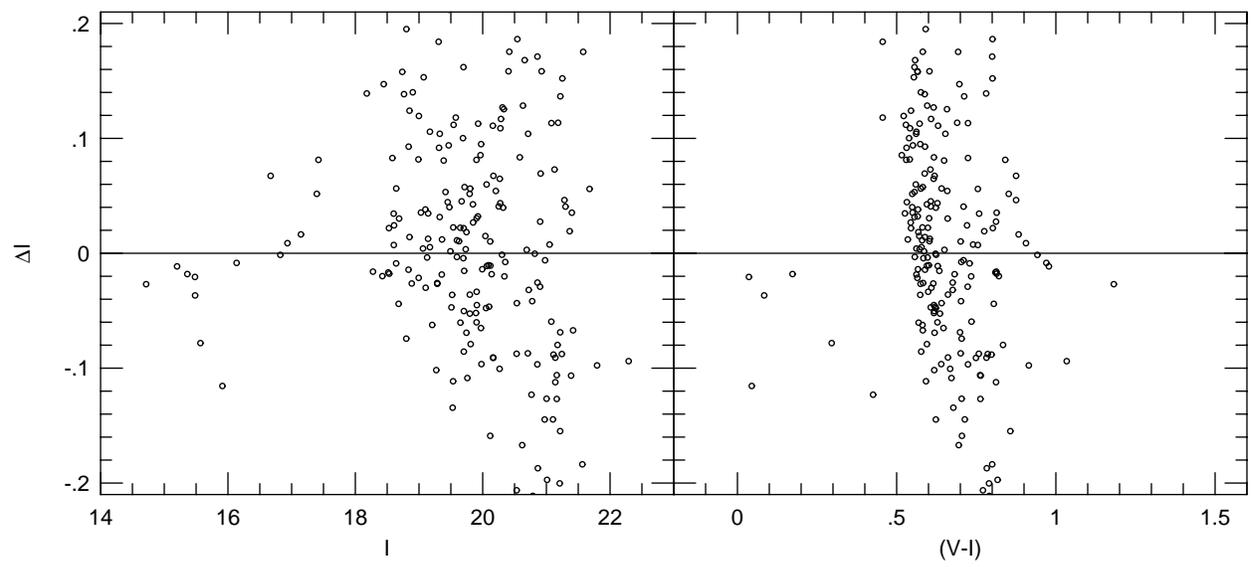
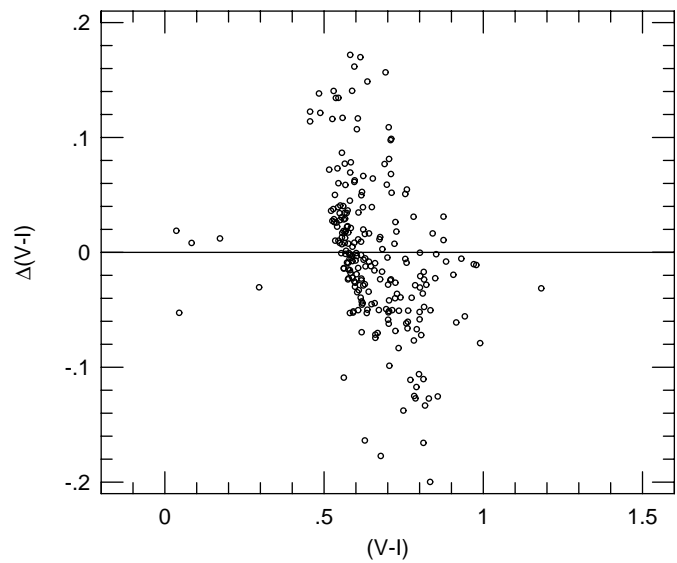
\end{document}